# Designing for effective heat transfer in a solid thermal energy storage system

Shomik Verma, Colin Kelsall, Kyle Buznitsky, Alina LaPotin, Ashwin Sandeep, Asegun Henry*

Department of Mechanical Engineering, Massachusetts Institute of Technology, 77 Massachusetts Avenue, Cambridge, MA 02139, U.S.A.
* ase@mit.edu

## 1. Abstract

Thermal energy storage using sensible heating of a solid storage medium is a potential low-cost technology for long-duration energy storage. To effectively get heat in and out of the solid material, channels of heat transfer fluid can be embedded within the storage material. Here we present design principles to improve performance of channel-embedded thermal energy storage systems, and we apply these principles to a high-temperature system using graphite as the storage material and liquid tin as the heat transfer fluid. We first analyze the impact of geometry and material properties on the performance of the system, determining the ideal channel spacing and length to achieve high (dis)charge temperature uniformity. We then analyze how controlling the fluid flowrate, heating infrastructure, and heat engine can increase discharge power uniformity and accelerate charging. Finally, we model 100 high-temperature graphite storage blocks using a porous media approximation and implement the developed design principles to demonstrate significant improvement in performance for both discharging (constant discharge power for >90% of rated duration) and charging (>90% charged within 4 hours). Overall, the hierarchical design procedure presented here enables the design of cheap yet high-performing solid thermal energy storage systems.

## 2. Introduction

Renewables such as solar produce clean but intermittent electricity. Large-scale penetration of solar requires long-duration energy storage (>10 hours) to reliably meet demand and provide cleanly generated electricity at night or on cloudy days.[1] Energy systems modeling shows the energy capacity cost (ECC, dollars per amount of thermal energy stored in kWh) is an important metric for evaluating long-duration energy storage technologies, and that an ECC of less than $20/kWh is critical for cost-competitiveness of renewables with storage technologies.[2]

Thermal energy storage (TES) is one promising technology with low ECC.[3] In this energy storage technology, electricity is stored in the form of heat. Electricity can be converted to heat through resistance heating or heat pumps, and heat can be stored through latent or sensible heating of a material surrounded by insulation to retain the heat, and heat can be converted back to electricity using a variety of heat engines including turbines, thermoelectrics, or thermophotovoltaics.[4,5]

There are a few important considerations for the storage material. First, we must choose cheap materials that can store high amounts of heat because of the emphasis on low ECC. Second, we should maximize the amount of thermal energy stored per volume to reduce insulation costs. Lastly, we should maximize the operating temperature to ensure high heat to electricity conversion efficiency.

As shown in Figure 1(a), the materials with the lowest cost per unit thermal energy, highest energy density, and highest operating temperature are generally solids and rely on sensible heating. These solid storage systems require a heat exchanger to be integrated into the storage material for charging or discharging.[6] There are different types of heat transfer mechanisms, including a radiative heater relying on radiation between a hot emitter and the storage medium, direct Joule heating of the storage medium, and a fluid-based heat exchanger using a heat transfer fluid (HTF) to carry the heat. Radiative heating can have high heat transfer resistance, and direct Joule heating depends on the electrical properties of the storage medium reducing flexibility in material choices. In contrast, fluid-based heat exchange enables low heat transfer resistance and cheap storage materials, so is often preferred.

The challenge of using a solid storage material is that it may perform poorly from a heat transfer perspective. Specifically, the system would ideally output a constant temperature (and power) with time, to operate like an electrochemical battery, be in alignment with how the grid is currently operated, help improve reliability, and take advantage of price spikes. Systems that use liquids as the storage material, in either one or two tanks, can generally output high-temperature heat as long as hot fluid remains in the tank.[7–10] Solid storage systems can be significantly worse, because as the storage material cools, the outlet temperature of the HTF decreases, potentially reducing the amount of power discharged from the TES system.[9] Despite being cheap and energy dense, this can put solid storage systems at a disadvantage when compared to other forms of energy storage, and it is therefore important to design the heat transfer infrastructure for maximum performance.

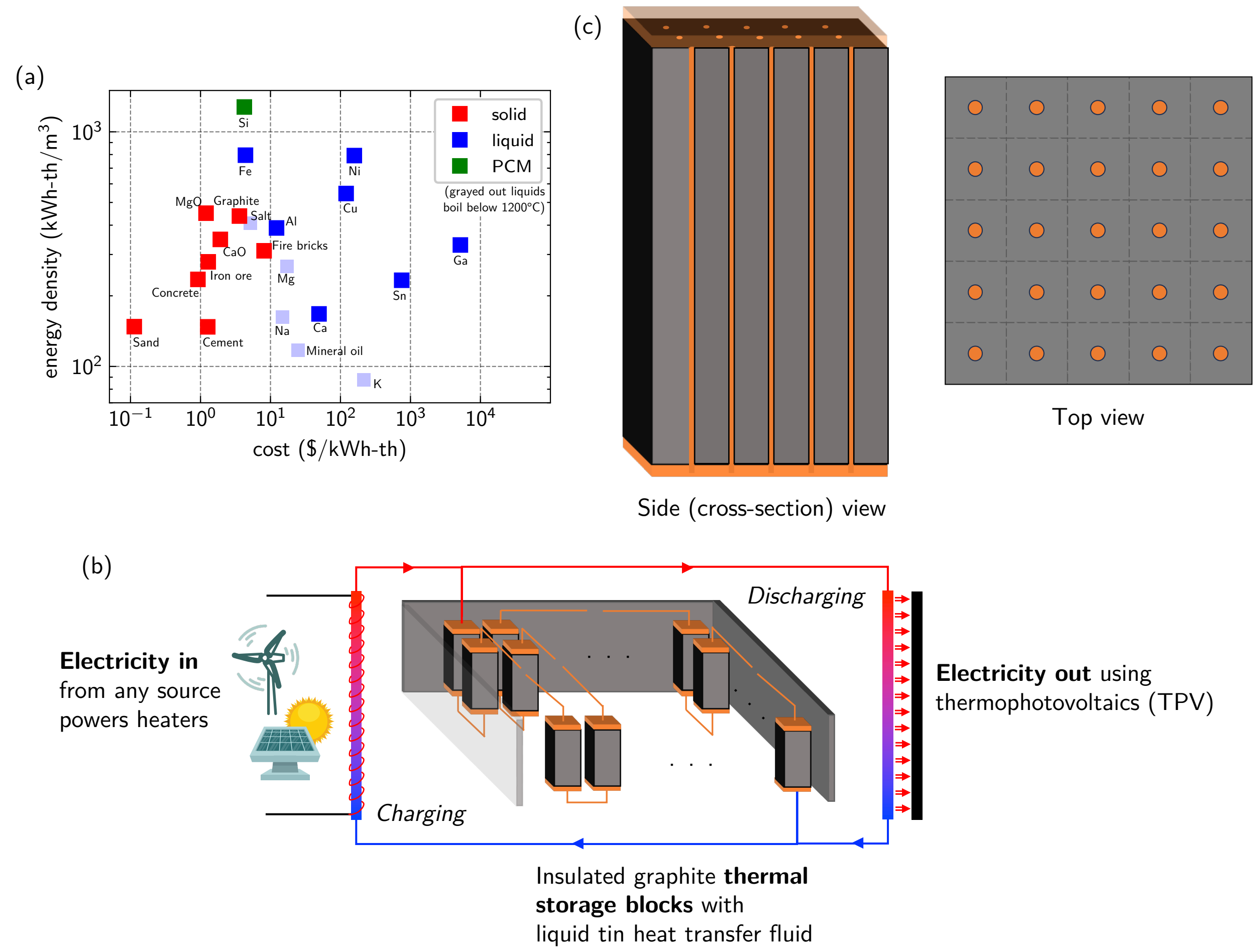

*Figure 1: (a) Plot of cost vs. energy density of a variety of storage materials, labeled by the phase which stores the maximum amount of heat (assuming 500°C for sensible heating) without decomposing. Solids enable the lowest cost and highest energy densities, as well as other benefits including achieving high temperatures and limited corrosion. Data used to create this plot is presented in Section S1. (b) Thermal energy grid storage (TEGS) system considered in this work. The system can be divided into charging and discharging infrastructure. For charging, electricity from any desired source is used to heat up liquid tin to high temperatures (~2400ºC) with resistive heating. The hot tin then flows through graphite storage blocks to heat them up, storing the energy in the form of sensible heat. During discharge, cold (~1900ºC) tin flows through the graphite blocks, heating up, and the hot tin then flows to the power block featuring multi-junction thermophotovoltaic (TPV) cells which convert the heat to electricity. Graphite is the storage medium and liquid tin is the heat transfer fluid. (c) Side cross-section view of a graphite storage block with embedded liquid tin channels, and top view of a graphite storage block showing the tin channel inlets. Dashed lines show how the larger block can be split into 25 sub-blocks, each with only 1 channel.*

Several previous works have investigated different heat exchanger designs and performance improvement strategies. One simple heat exchanger design is to use particles of solid storage medium surrounded by a heat transfer fluid, as investigated by many researchers including ELSihy et al.,[11] Rodriguez and Lemos,[12,13] and Marti et al.,[14] who also discuss performance improvements through materials and operational changes. However, this geometry relies on the storage material being available as particles, and often features high pressure drops in the fluid.[15] For other heat exchanger designs, Wu et al. compared four different geometries, including a block with channels of HTF drilled into it (channel-embedded block), solid rectangular plates with HTF flowing around them (parallel-plates), a bundle of solid rods with HFT flowing around it (rod-bundle), and the packed bed configuration as discussed previously.[16] They found that the packed-bed configuration had the best outlet temperature uniformity, indicating optimization needs to be done for the other configurations.[16] Among the other configurations, the channel-embedded block allows the most flexibility in the form of the storage medium, as channels for HTF flow simply need to be drilled within or placed around the storage material, instead of having to form the storage block into different shapes. Therefore, the channel-embedded block configuration is investigated in this work.

Some previous studies have investigated how best to optimize these channel-embedded solid systems for charging or discharging performance, using various parameters. Jian et al. investigated the effects of diameter ratio, channel length, flowrate, and number of tubes on discharge performance, optimizing storage cost based on an analytical solution using a corrected lumped capacitance model.[17] There have also been experimental validations of this lumped capacitance model.[18,19] Prasad et al. investigated the effect of the number of tubes and flowrate on charge and discharge performance.[20]

While the studies above only considered a single block, large-scale TES systems will consist of many blocks operating in an interconnected way. Doretti et al. considered series vs. parallel arrangements of blocks, as well as the diabatic case of heat loss to the environment, finding the optimal configuration using the minimal number of blocks.[21] Kuang et al. investigated the effect of increasing the flowrate with time to counterbalance the decreasing temperature in two series-connected rod-bundles, enabling the extension of the stable discharging time by 1.6 times.[22]

Despite these previous works, there are still significant gaps in understanding exactly what is required for effective design. Many works consider only charging or discharging in isolation but not how designing for one condition impacts the other. Previous works often use a corrected lumped-capacitance approximation, which may not be valid for the large Biot numbers associated with some HTFs (such as liquid metals), and which ignores axial conduction. Further, previous works often only choose one storage material, and the impact of varying material properties on performance is not made clear. Perhaps most importantly, no previous work has considered a multi-block configuration of the size required for grid-scale deployment. This configuration requires analysis of optimal flow distribution and heat transfer between the blocks, which is particularly relevant at high temperatures (>1500ºC).

Therefore, this study will focus on how to design for effective heat transfer in a large-scale, channel-embedded solid thermal energy storage system during both discharging and charging. We have developed general design principles to maximize performance, with performance metrics defined for constant-power discharge and accelerated charging. We then use these design principles to analyze a specific solid sensible thermal energy grid storage system called TEGS (Thermal Energy Grid Storage). While only the TEGS system is analyzed here, the design principles are applicable to many solid storage systems, as discussed throughout.

An overview of the TEGS system is shown in Figure 1(b). In this system, graphite is used as the storage medium, and can be found in the form of large channel-embedded blocks (as shown in Figure 1(c)) which may be connected in series or parallel. Liquid tin is used as the HTF. During charging, resistance heaters are used as the heat input from electricity, and during discharging, thermophotovoltaics (TPV) are the heat engine. TPV works by converting the light radiated by a heat source into electricity. The system operates from a minimum temperature of 1900°C to a maximum of 2400°C, to maximize TPV efficiency and power density.[23]

There are many questions we would like our design principles to answer. How big and how many channels should there be per block? How do the storage blocks best interface with the charging and discharging infrastructure? How should the HTF flow between the blocks be configured? What is the best arrangement of blocks considering radiation between them?

One way to answer these questions would be to model the entire storage system, including coupled fluid flow and heat transfer, and radiation between the blocks. However, this quickly becomes computationally intractable due to the large size (~1GWh) and long duration (~30 hours) of the systems of interest. Therefore, we have developed a hierarchical design and modeling methodology to determine design principles that result in optimal performance. We started with analyzing the impact of physical (material properties, geometry) and operational (flowrate) variables on the performance of a single HTF flow path, then apply the insights generated to a model of a multi-block grid-scale TES system.

## 3. Theory and preliminary calculations

Given a storage system, the duration of storage (hours) is defined as the thermal energy (Wh) divided by the power extracted (W). Thermal energy is defined as:

$$E = m_s \int_{T_{min,s}}^{T_{max,s}} c_{p,s}(T)dT \approx m_s c_{p,s} \Delta T_s$$

where $m_s$ is the total mass of the solid blocks, $c_{p,s}(T)$ is the specific heat of the solid as a function of temperature, and $T_{max,s}$ is the maximum operating temperature while $T_{min,s}$ is the minimum operating temperature of the storage system. If the specific heat does not change significantly over the temperature range, it can be considered constant and the integral can be simplified, where $\Delta T_s = (T_{max,s} - T_{min,s})$. The thermal power extracted (or provided) is calculated by:

$$Q = \dot{m}_f [c_{p,f}(T_{f,out}) \cdot T_{f,out} - c_{p,f}(T_{f,in}) \cdot T_{f,in}] \approx \dot{m}_f c_{p,f} \Delta T_f \quad (1)$$

where $\dot{m}_f$ is the flowrate of HTF through the system, $c_{p,f}(T)$ is the specific heat of the HTF as a function of temperature, $T_{f,out}$ is the HTF outlet temperature from the storage blocks, and $T_{f,in}$ is the HTF inlet temperature to the storage blocks. Again, if the specific heat can be considered constant, this expression can be simplified where $\Delta T_f$ is the difference in HTF outlet and inlet temperatures.

During discharge, hot HTF leaves the storage heat exchanger, is cooled by the heat engine, and returns to the storage system cooled (top right loop in Figure 1(b)). The temperature of cold HTF entering the storage $(T_{f,in})$ is assumed to be the minimum operating temperature of the system $(T_{min,s})$, and ideally the storage would heat up the HTF such that its outlet temperature $(T_{f,out})$ is the highest operating temperature of the system $\left(T_{max,s}\right)$ for the entire discharge. This is because providing the maximum temperature enables higher heat engine efficiency, improving the round-trip efficiency of the thermal storage system. Constant temperature is optimal because, as mentioned previously, providing a constant power output enables operation like an electrochemical battery, improving reliability, maximizing revenue, and easing grid integration. Conversely, during charging, cold HTF leaves the storage system, is heated by the charging infrastructure, and hot HTF returns to the storage system (bottom left loop in Figure 1(b)). The temperature of the hot HTF entering the storage $\left(T_{f,in}\right)$ is assumed to be the maximum operating temperature of the system $\left(T_{max,s}\right)$, and ideally the HTF would transfer all its heat to the storage medium such that its outlet temperature $\left(T_{f,out}\right)$ from the storage blocks is the lowest operating temperature $\left(T_{min,s}\right)$. Thus, $\max\left(\Delta T_f\right) = T_{max,s} - T_{min,s} = \Delta T_s$ and $Q_{max} = \dot{m}_f c_{p,f} \Delta T_s$.

We can now define a characteristic time scale $\tau$, non-dimensional time $t^*$, non-dimensional temperature as a function of time $\Theta(t)$, and non-dimensional temperature as a function of non-dimensional time $\Theta^*(t^*)$, for a given storage system:

$$\tau = \frac{E}{Q_{max}} = \frac{m_s c_{p,s}}{\dot{m}_f c_{p,f}} \tag{2}$$

$$t^* = \frac{t}{\tau}$$

$$\Theta(t) = \begin{cases} \dfrac{T(t) - T_{min}}{T_{max} - T_{min}} & during\ discharge \\ \dfrac{T_{max} - T(t)}{T_{max} - T_{min}} & during\ charge \end{cases}$$

$$\Theta^*(t^*) = \Theta(t)$$

In reality, the ideal outlet temperatures discussed above are not achievable. As a solid thermal storage medium discharges and cools, the outlet temperature of the HTF decreases. Conversely, during charging the HTF outlet temperature increases as the storage medium heats up. Thus, there is non-ideality in real system operation. To get some intuition for this phenomenon, we can simulate a real system as an example. We model a simplified version of the TEGS system (a single block of graphite as the solid storage medium and one flow channel liquid tin as the HTF), with the specific geometric, material, and operational parameters described in Section 5.1. In this simulation, we start with a graphite block at $\Theta = 1$ and flow tin in at $\Theta = 0$ until the entire block is at $\Theta = 0$. Figure 2 shows the idealized vs. realistic non-dimensional temperature profiles for tin outlet temperature during (dis)charge, for the system described above with different graphite thermal conductivities, to show the impact of material properties on performance. From this example, we see that while the higher thermal conductivity cases start to approach the ideal limit, there is still significant room for improvement.

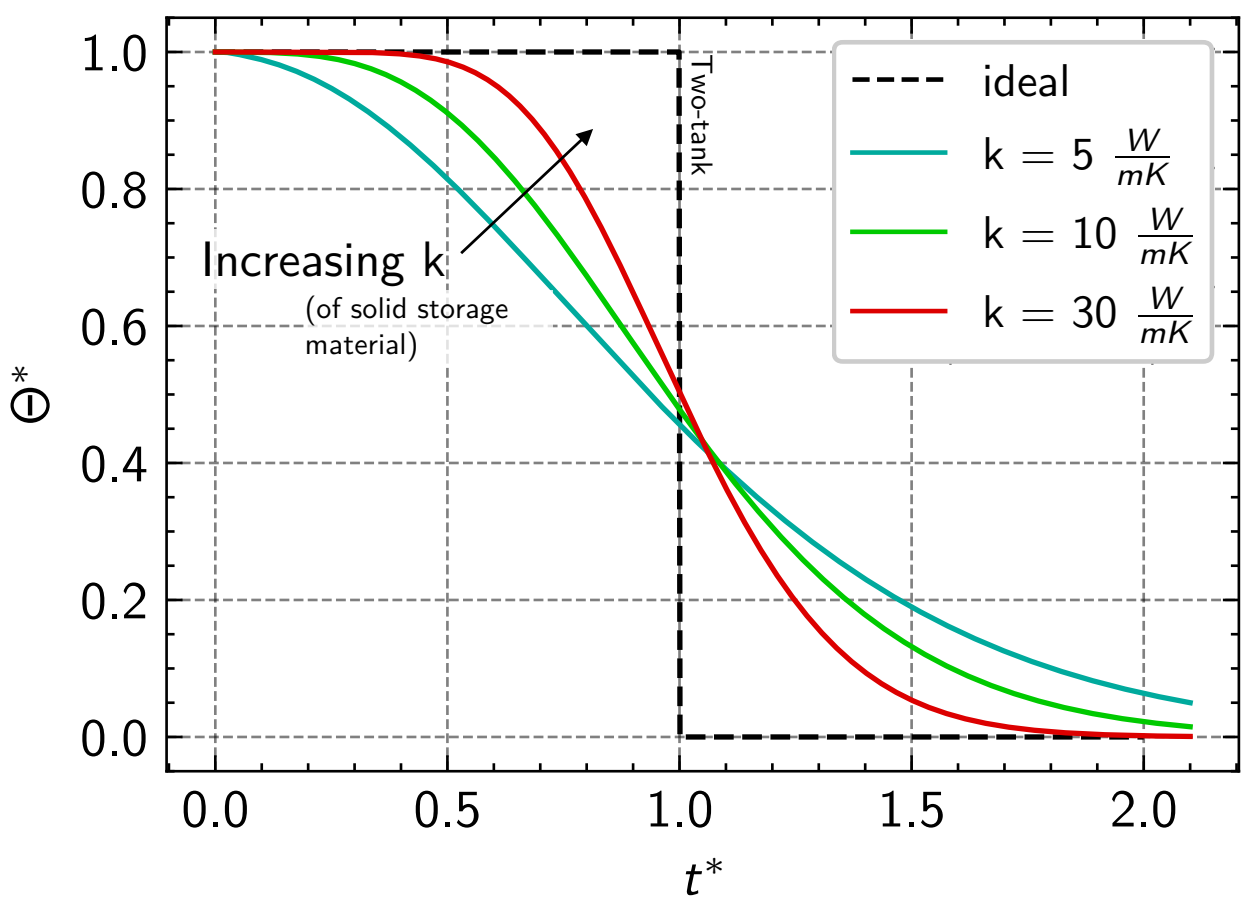

*Figure 2: Comparison of non-dimensional tin outlet temperature for the ideal (dis)charge profile of a two-tank thermal storage system or electrochemical battery (dashed line), and the non-ideal (dis)charge profiles of channel-embedded passive system (solid lines, the nondimensional (dis)charge temperature profiles as a function of non-dimensional time for the 3 thermal conductivities shown). The block is initially at Θ=1 everywhere, and tin enters at Θ=0 until the block is at Θ=0 everywhere. Note that the same curves are obtained using either the discharging or charging definition of $\Theta$.*

One simple strategy for improvement is over-sizing storage capacity such that the storage system is never discharged more than e.g., 25%. However, this directly increases costs, as large amounts of storage material are never fully utilized. It also reduces the charging speed as the system would always be at a high temperature, reducing the $\Delta T$ available for charging power. These effects are discussed further in Section S2. Because of these issues, oversizing is ruled out as a potential solution.

Instead, we are interested in determining how performance can be maximized based on changes to geometry, operation, and block arrangement in the storage system. We first define a temperature-based figure of merit (FOM) as,

$$FOM_T = \int_0^1 \Theta^*(t^*)\,dt^*$$

If $\Theta^*(t^*)$ was an idealized square curve, this integral would evaluate to 1, so this FOM serves as an indicator of how ideal the actual (dis)charge profile is. While this is a useful non-dimensional metric, it has some limitations, namely that it only considers temperature and not power. What we care about practically are (a) how long we can output a constant electrical power during discharge, and (b) how fast the system can charge. Thus, to be more explicit, we first define electrical discharge and charging power as

$$P_{out} = Q_{out}\eta_{discharge}$$

$$Q_{in} = P_{in}\eta_{charge}$$

where $\eta_{discharge}$ is the efficiency of the heat engine used to generate electricity $(P_{out})$ from the thermal power discharge from the storage $(Q_{out})$, and $\eta_{charge}$ is the efficiency of the heaters used to generate heat $(Q_{in})$ from electricity $(P_{in})$. We then define two power-based FOMs. For discharge,

$$FOM_{P,discharge} = \frac{t_{P_{max}}}{\tau}$$

which compares the amount of time that the storage system can output the nominal electrical power $\left(t_{P,max}\right)$, to the rated storage duration $(\tau)$ as defined in Equation 2. We define another FOM for charging,

$$FOM_{P,charge} = \frac{\int_0^\tau \dot{m}(t) c_{p,Sn} \Delta T(t)\, dt}{\eta_{charge} E}$$

which compares the integrated electrical power input over the charging duration to the energy capacity, and is essentially the state of charge of the system after the charging duration. In this work, we seek to maximize these FOMs.

Because modeling the entire storage system is computationally intractable due to the large system size (~1 GWh), long durations (~10 hours), and large range of geometry sizes (~1mm to ~1m), we propose a hierarchical design procedure to methodologically model and optimize performance, in increasing levels of fidelity to a real system, as summarized in Figure 3. First, as shown in Figure 1(c), each large block can be separated into individual sub-blocks with only 1 channel embedded. Even with this simplified geometry, there are many variables (as shown in Figure 3(a)), so we use dimensional analysis to understand the state space and evaluate the effects of physical (geometric and material) parameters on $FOM_T$. In addition to the physical variables, operational variables such as flowrate and heat engine operation are also important. Thus, second, as shown in Figure 3(b), we simulate a single-channel cylinder during both discharging and charging to determine the impact of operational variables such as flowrate, heater power, and heat engine control on $FOM_{P,discharge}$ and $FOM_{P,charge}$. Finally, as shown in Figure 3(c), we model a large-scale storage system with multi-channel blocks using a porous media approximation to determine the effects of block interconnection, block arrangement, and radiation between the blocks on the power FOMs.

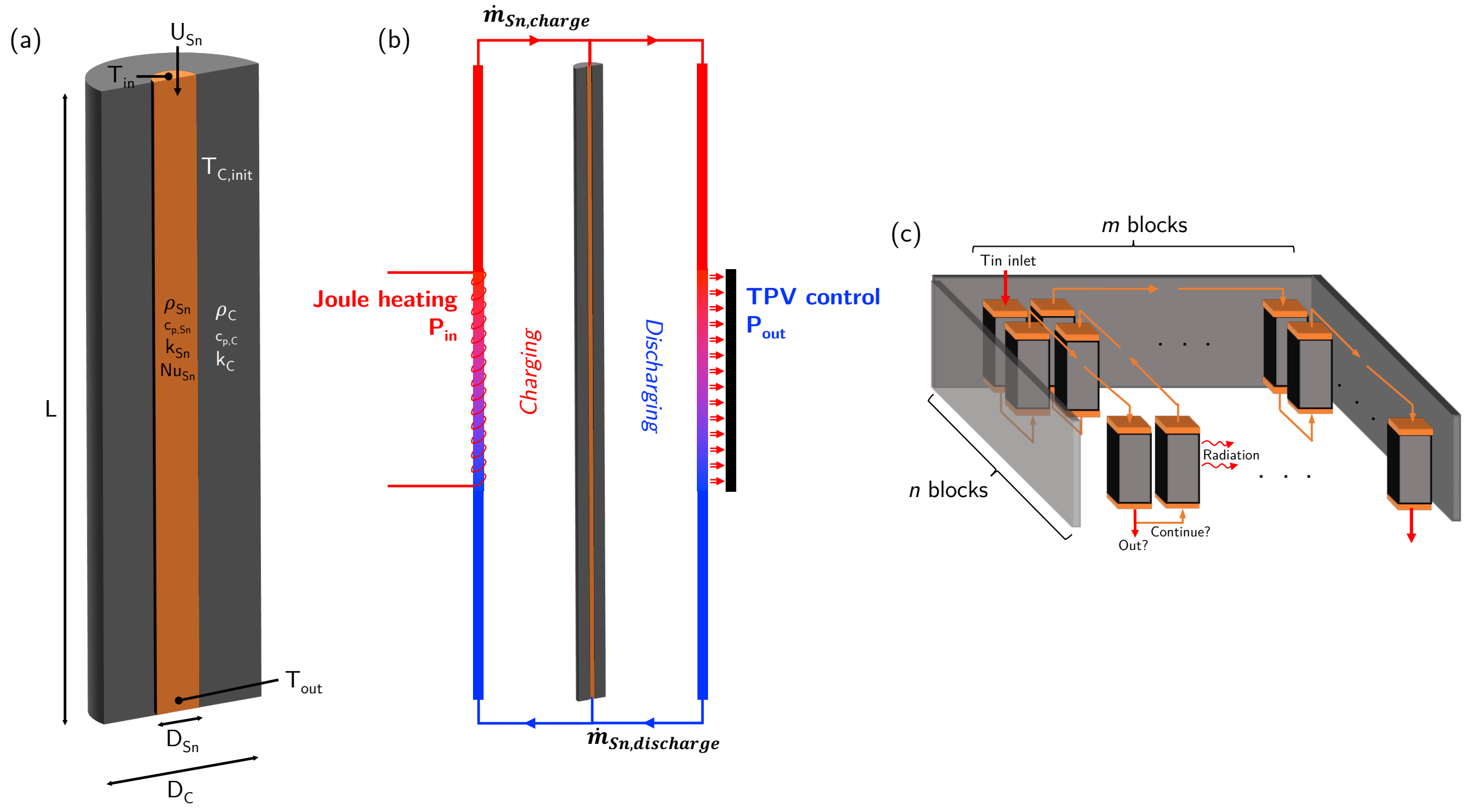

*Figure 3: Schematic of the three phases of the hierarchical design procedure presented here. (a) Geometry optimization of a block with a single channel by varying the physical parameters, (b) operational optimization by varying flowrate and controlling the heater power and heat engine, and (c) large-scale model with multiple large blocks to understand the impact of flow configuration (how many blocks should the tin flow through before exiting), block arrangement (what is the impact of n and m on performance), and radiation (how does the high temperatures of the blocks impact performance).*

# 4. Results and discussion

## 4.1. Impact of storage medium physical parameters on performance

First, we utilize dimensional analysis to help determine the impact of geometric parameters and material properties on performance. We are particularly interested in (a) how to design the channel-embedded heat exchanger, namely the channel spacing and length, and (b) how the thermal properties of the storage medium affect performance. For this analysis, we simplify the geometry to consider a cylindrical storage block with a single HTF channel. Figure 3(a) shows this simplified geometry along with key parameters in the analysis. We consider a cylindrical HTF channel of diameter $D_f$ surrounded by solid with outer diameter $D_s$, both having length L. The HTF flows at velocity $U(t)$ and has a convective heat transfer coefficient with the solid of $h$. Material properties such as density, specific heat, and thermal conductivity of both fluid and solid are also needed.

Conducting dimensional analysis on the 15 variables with 4 dimensions results in 11 dimensionless $\Pi$ groups, all of which are provided in Section S3. Rearranging in terms of known variables, we find that

$$FOM_T = f\left(\frac{L}{D_s}, \frac{D_s}{D_f}, \frac{k_s}{k_f}, \frac{c_{p,s}}{c_{p,f}}, \frac{\rho_s}{\rho_f}, \frac{UD_f}{\alpha_s}, \frac{hD_f}{k_f} = Nu_{D_f}, \frac{\rho_f U D_f}{\mu_f} = Re_{D_f}, \frac{\mu U^2}{k_f \Delta T} = Br\right)$$

These 9 dimensionless groups define the state space of possible channel-embedded thermal storage designs. By varying these 9 parameters, we can explore the impact of different variables on the FOM.

To demonstrate how this tool can be used in a real system, we consider the TEGS system with liquid tin as the HTF and graphite as the storage medium. From previous works,[24] we know that the densities and specific heats of different grades of graphite do not vary significantly, while thermal conductivity does. Tin material parameters are considered constant in this analysis. Further, we assume the volume ratio of tin to graphite to be 1%, which enables favorable technoeconomics as the cost of tin is ~30 times higher than that of graphite, and higher volume fractions increase the cost per energy of the system.[25] This 1% volume fraction implies $\frac{\pi D_s^2 - \pi D_f^2}{\pi D_f^2} = 100 \Rightarrow \frac{D_s}{D_f} = \sqrt{101} \approx 10$ is a constant. Additionally, we assume laminar tin flow meaning constant $Nu = \frac{hD_f}{k_f}$. This assumption is validated in Section S3.

Based on these assumptions, we can simplify the expression above such that $FOM_T$ can be written as a function of 3 dimensional groups:

$$FOM_T = f\left(\frac{L}{D_s}, \frac{UD_f}{\alpha_s}, \frac{k_s}{k_f}\right)$$

Where $\frac{L}{D_s}$ is an aspect ratio, $\frac{UD_f}{\alpha_s}$ compares tin advection to graphite conduction, and $\frac{k_s}{k_f}$ compares the thermal conductivities of the two materials. We now have a more manageable 3-dimensional state space. We can use a finite element model (as discussed in Section 5.1) of the simplified geometry to calculate values of $FOM_T$ for points in this 3D state space, and determine where it is optimal to operate. 2D slices of the 3D state space are shown in Figure 4(a), which shows $FOM_T$ vs $\frac{L}{D_s}$ and $\frac{UD_f}{\alpha_s}$ for various values of $\frac{k_s}{k_f}$. Visually, we first note that we can quickly identify the region of design space we want to operate in, namely the high-FOM yellow region. This allows quick evaluation of whether a designed configuration will be high performing. However, we also note that these plots look very similar across different $k$ ratios, which is unexpected because we previously determined a lower solid thermal conductivity $k_s$ should reduce performance. Instead, we notice that the impact of $k_s$ comes in the x-axis, since $k_s$ also appears in $\alpha_s = \frac{k_s}{\rho_s c_{p,s}}$. Thus, for a given flow geometry, lower values of $k_s$ shift the operating point to the right, away from the high-FOM yellow region, as $\frac{UD_f}{\alpha_s}$ increases.

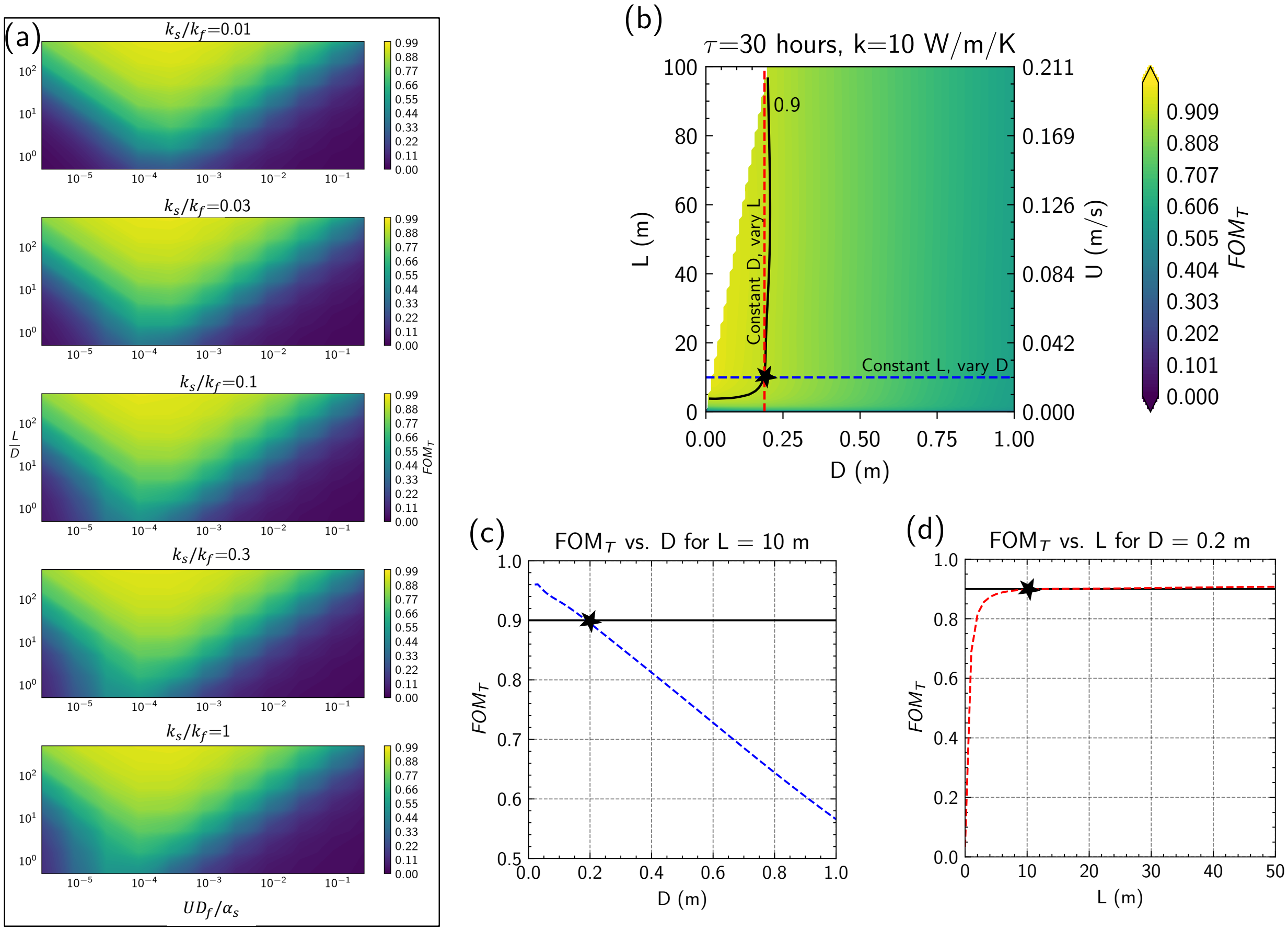

*Figure 4: (a) Results from dimensional analysis showing the complete parameter space of the channel-embedded thermal energy storage system called TEGS, showing 2D slices of $\frac{L}{D}$ and $\frac{UD}{\alpha}$ vs. $FOM_T$ for constant $k_s/k_f$. (b) Dimensioned sweeps of the parameter space for a given graphite thermal conductivity (10 W/m/K) and storage duration of 30 hours. The solid line is an isocontour of FOM=0.9, and the star shows the optimal operating point for largest diameter and smallest length to achieve high FOM. (c) depicts a line of FOM vs. D corresponding to the blue dashed line in (b). (d) and (e) are equivalent plots for a storage duration of 10 hours.*

While this non-dimensional map is useful, it is not immediately clear how to use the map to design a high-performing system. We thus demonstrate an example of how the mapped design space can be used practically, to design the geometric parameters of the TEGS system. We first consider a storage duration ($\tau$) of 30 hours and a graphite thermal conductivity ($k_s$) of 10 W/m/K. We then choose a length $L$ and graphite diameter $D_s$ (which also defines $D_f$ since $\frac{D_s}{D_f}$ was fixed at 10) of the storage block. We can then use the duration $\tau$ and length $L$ to calculate the flow velocity $U$, by re-writing Equation 2 as

$$\tau = \frac{\rho_s V_s c_{p_s}}{\dot{m}_f c_{p_f}} = \frac{\rho_s L\pi({D_s}^2 - D_f^2)c_{p_s}}{\rho_f U\pi D_f^2 c_{p_f}} = \frac{\rho_s L\left[\left(\frac{D_s}{D_f}\right)^2 - 1\right]c_{p_s}}{\rho_f U c_{p_f}} = 100\frac{\rho_s}{\rho_f}\frac{c_{p,s}}{c_{p,f}}\frac{L}{U}$$

$$\Rightarrow U = \left(\frac{100\rho_s c_{p,s}}{\rho_f c_{p,f}}\right)\left(\frac{L}{\tau}\right)$$

Thus, using the input variables $\tau, k_s, L,$ and $D_s$, we can calculate $U$ and therefore all 3 of the dimensionless groups, and can use the non-dimensional map to determine the $FOM_T$. The results are shown in Figure 4(b). A few trends are evident. First, reducing the diameter of the graphite $D_s$ (and therefore the tin channel $D_f$) improves $FOM_T$, as seen in Figure 4(c). This can be explained by the shorter conduction distance in the graphite, so the heat can more easily enter/leave the graphite, improving its performance. However, if we were to only reduce the graphite diameter $D_s$ while keeping $D_f$ constant, this would hurt the energy density (energy per volume) and cost of the system, since the tin HTF would take up more volume. Instead, because we scale down the fluid diameter $D_f$ with $D_s$, we retain the energy density and low cost of the system while improving performance.

Although the FOM monotonically increases with reduced $D_f$, the reduction in $D_f$ does cause an increase in pressure drop, which should be avoided due to higher parasitic power consumption of the motor, wear on mechanical parts, and higher leakage potential. We recall that pressure drop in a tube is defined as:

$$\varDelta P = \frac{\rho_{Sn}U^2 L}{2D_f}f(Re) \Rightarrow \varDelta P_{lam} \propto \frac{\mu_{Sn}UL}{D_f^2}, \Delta P_{turb,rough} \propto \frac{\rho_{Sn}U^2 L}{2D_f}$$

suggesting that we should not go to the extreme of $D_f \to 0$ if we want to prevent high pressure drops. This demonstrates the tradeoff between low $D_f$ improving performance while increasing pressure drop.

The second important trend is that increasing the tin channel length $L$ improves FOM, but this effect plateaus quickly, as seen in Figure 4(d). For the configuration considered here, the FOM stops improving after around 10m, suggesting that e.g. a 20m long channel-embedded block has equivalent thermal performance to 2 10m long blocks. This is important because, from the pressure drop definition above, it is beneficial to minimize L. We can therefore conclude that placing $n$ 10m long pipes in parallel has equivalent thermal performance to a single $10n$ m long pipe, and the $n$ pipes in parallel have $4n$ (laminar) to $16n$ (turbulent rough) times lower pressure drop.

Thus, optimizing the FOM with the constraint of minimizing pressure drop (minimizing L and maximizing D) suggests an optimal operating point of $D_S = 20cm$ ($D_f = 2cm$) and $L = 10m$ to achieve an $FOM_T$ of 90%.

This analysis is repeated for storage duration $\tau = 100$ hours and graphite thermal conductivity $k_s = 1$ W/m/K in Section S4 to demonstrate applicability to a different configuration.

While only the TEGS system is considered here, the design procedure can be applied to any channel-embedded solid thermal storage system, by (a) applying the dimensional analysis derived here for the system in question and mapping the state space, (b) selecting a storage duration and material properties, and (c) determining the maximum diameter and minimum length that provides a high FOM (e.g. 0.9) based on the state space map.

Now that an understanding of the effect of physical parameters has been developed, we can consider operational variables and how the storage block would be integrated with charge and discharge infrastructure.

## *4.2. Coupling the storage medium with (dis)charging infrastructure*

Given a solid storage geometry from the previous section, we can now determine how the operation of the storage system impacts performance. There are two specific items to improve. First, output electrical power should ideally be constant over time, at the rated output power of the storage system, which is what utilities would expect and is what power plants and electrochemical batteries provide. However, as seen in Figure 2, temperature slowly decreases over time, potentially reducing power output. Second, since the intent of the storage system is to improve the reliability of variable renewable energy sources, the system should be able to charge quickly when renewable energy is available; it is anticipated that there will be greater value on the grid for a storage system that can charge faster than it discharges.[26,27] Thus, the two reasons for trying to optimize operational variables are (a) achieving closer to constant electrical power output on discharge and (b) faster charging.

### 4.2.1. Constant power output during discharge

For discharging, while we could use $FOM_T$ above, it is not fully appropriate for a few reasons. First, it only considers temperature and not power directly. Second, the setup of $FOM_T$ assumes the HTF flowrate does not change with time, which is the simplest operation mode but may not be optimal. Third, the integral form obscures potentially important discharge dynamics – for example, a system that outputs 100% of rated power for 80% of rated duration has the same $FOM_T$ as a system that outputs 80% of rated power for 100% of rated duration. However, the first system is more ideal because utilities expect the rated power output. Of course, the most ideal system would output 100% of rated power for 100% of rated duration; therefore, we define an FOM directly based on duration of constant power output, namely,

$$FOM_{P,discharge} = \frac{t_{P_{max}}}{\tau}$$

where $t_{P,max}$ is the amount of time that the storage system can output the nominal electrical power, and $\tau$ is the storage duration as in Equation 2.

We now introduce a technique to maximize $FOM_{P,discharge}$ for a general thermal storage system. First, if we want to output constant electrical power, we need to ensure constant thermal power is being output. We noted previously that the cause of thermal power decrease over time is the drop in outlet temperature due to the storage medium being cooled. To counteract this, as the outlet temperature (and therefore $\Delta T_f$) drops, we could concurrently ramp up mass flowrate of the HTF, keeping output thermal power constant as defined by $Q(t) = \dot{m}_f(t) c_{p,f} \Delta T_f(t)$. Thus, we have

$$\dot{m}_f(t) = \frac{Q_{max}}{c_{p,f} \Delta T_f(t)}$$

We must bound $\dot{m}_f$ however, to avoid extreme flowrates as $\Delta T_f$ drops to 0. Therefore, we define a maximum flowrate $\dot{m}_{max}$ and a maximum flowrate factor $f = \frac{\dot{m}_{max}}{\dot{m}_f(t=0)}$. Thus, the flowrate is able to keep up with the drop in $\Delta T_f$ until $\dot{m}_f(t) = \dot{m}_{max}$, and we can calculate the $\Delta T_f$ at that point as

$$\Delta T_{f,thres} = \frac{Q_{max}}{c_{p,f} \dot{m}_{max}}$$

where we have defined a threshold $\Delta T_{f,thres}$ below which output power starts to decrease because the flowrate cannot increase further. This suggests that the thermal power output as a function of time is

$$Q_{out}(t) = \begin{cases} Q_{max} & when\ \Delta T_f(t) > \Delta T_{f,thres} \\ \dot{m}_{max} c_{p,f} \Delta T_f\ (t) & when\ \Delta T_f(t) < \Delta T_{f,thres} \end{cases}$$

Higher values of $\dot{m}_{max}$ (and therefore higher $f$) will result in lower values of $\Delta T_{f,thres}$, from the equation above. This then implies that higher $f$ results in longer durations of constant power output at $Q_{max}$, since $\Delta T_f(t)$ will be higher than $\Delta T_{f,thres}$ for longer.

To make this approach clear, we consider an example using the TEGS system, of a single cylindrical graphite block of length $L = 10m$ and diameter $D_C = 0.2m$ (tin channel diameter $D_{Sn} = 0.02m$), with a graphite thermal conductivity of 10 W/m/K, based on the geometric optimization done in the previous section. We then consider a storage duration of 31.6 hours and a maximum flowrate factor $f = \frac{\dot{m}_{max}}{\dot{m}_{nom}} = 5.6$ (for demonstration purposes – we will consider additional values later). We can then numerically model the system while incorporating the approach above, with the results shown in Figure 5(a)-(c). As seen, as the outlet temperature of the tin HTF decreases (Figure 5(a)), we can increase the mass flowrate until a value of 5.6 times the flowrate at t=0 (Figure 5(b)). We can therefore keep the thermal power output constant for time $t_{P,max}$ (blue line in Figure 5(c)).

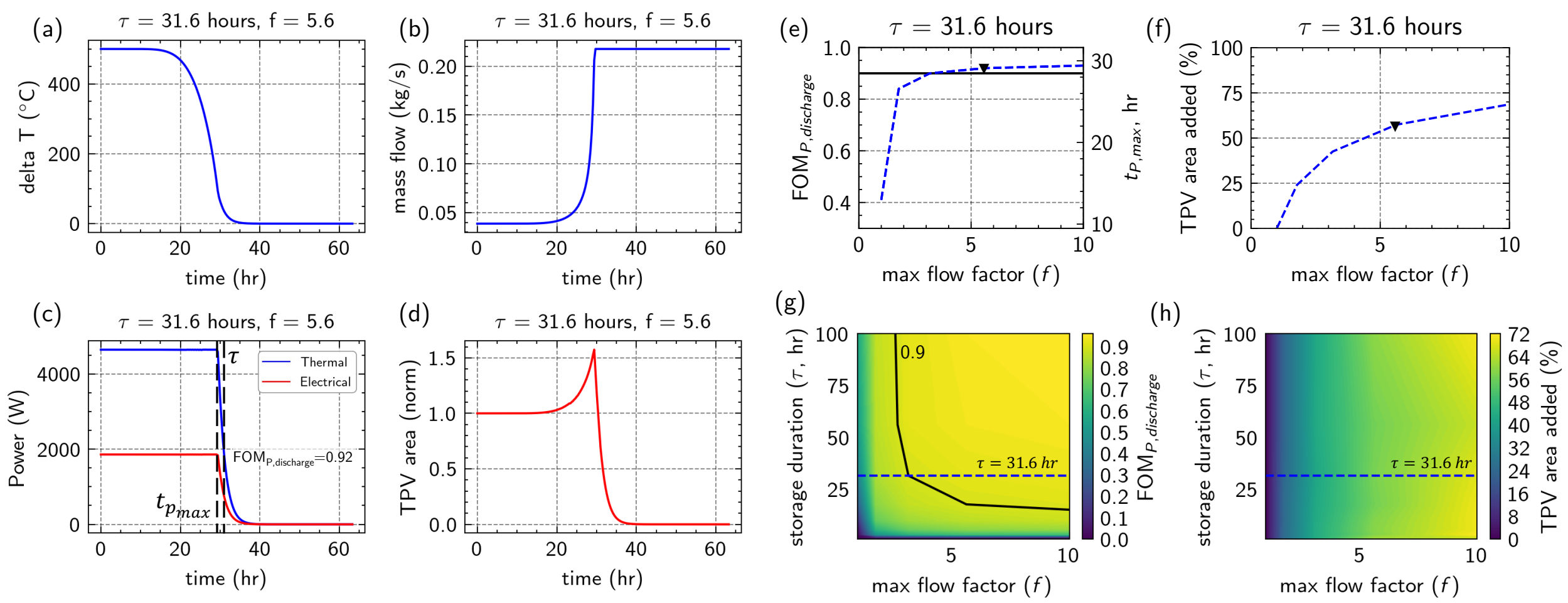

*Figure 5: Demonstration of the operational principle of flow ramping, to counteract the lower temperature tin outlet of the storage medium. First considering a discharge duration of 31.6 hours and a flowrate factor of 5.6, (a) Tin outlet temperature from the storage medium, subtracted by the inlet temperature (1900ºC), (b) flowrate of tin as a function of time showing the ramping, (c) thermal power of the tin output from the storage medium (blue line) and electrical power output from the TPV (red line), (d) Normalized TPV area required to convert the thermal power to electrical power at 40% efficiency, (e) the impact of allowing the flowrate to increase to a certain factor on the discharging FOM, for* $\tau = 31.6$*, with the triangle showing the f=5.6 case, (f) The percent of additional TPV area required to achieve the* $FOM_{P,discharge}$ *shown in (e), with the triangle showing the f=5.6 case. (g) Discharge FOM for various storage durations and flowrate factors, with solid line an isocontour of* $FOM_{P,discharge}$*=0.9, and the dashed line showing where* $\tau = 31.6$ *lies. (h) Percent of additional TPV area required to achieve the* $FOM_{P,discharge}$ *in* $(g)$*.*

While the approach above keeps output thermal power constant, we ultimately want to convert the thermal power to electrical power. Keeping electrical power constant is challenging, because as the discharging process progresses, the heat carried by the HTF becomes lower grade (lower temperature) and higher flowrate. The heat engine will likely have a harder time converting the lower grade heat to electricity. This suggests that the heat engine may need to be throttled at the beginning of discharge, meaning it would output less than its rated (maximum) power when temperature is maximum.

To give an example of how a heat engine may follow this strategy, we consider the heat engine used in the example TEGS system, which is thermophotovoltaics (TPV). TPV converts the energy radiated by a heat source, in this case the liquid tin HTF, to electricity. However, as the temperature of the HTF drops, the intensity of emitted radiation decreases, meaning more TPV area is required to keep the generated electrical power constant. To demonstrate this, we develop a simple model:

$$Q_{out} = q"\left(T_{f,avg}(t)\right) A_{TPV}(t) = \frac{P_{out}}{\eta_{TPV}}$$

where $q"(T)$ is the net (radiative) heat flux from the HTF to the TPV (modeled based on experimental data), $T_{f,avg}(t)$ is the 4th order average of HTF temperature as a function of time, and $A_{TPV}(t)$ is the active area of the TPV as a function of time. As the HTF temperature decreases, $q"(T_{f,avg})$ will also decrease, suggesting more TPV area will need to be exposed to the HTF to extract the same amount of thermal power from the tin ($Q_{out}$) and generate the same electrical power (the thermal power $Q_{out}$ is converted to electrical power $P_{out}$ at a TPV efficiency $\eta_{TPV}$). In the schematic of the TEGS system shown in Figure 1(b), the TPVs are fully facing the liquid tin, but this could be changed with time – this analysis suggests the TPV stick should be initially retracted, then as the temperature of the HTF drops, the TPV stick should be inserted further to expose more TPV area to the HTF. This is demonstrated in Figure 5(d), where we see the TPV area exposed to the tin HTF increases with time to keep the electrical power generated constant (Figure 5(c), red line).

The example above only considered a single maximum flowrate factor $f$, so next we consider multiple different values. As discussed previously, the benefit of increasing $f$ is longer duration of constant discharge power as the higher flowrate can counteract the lower HTF outlet temperature. However, the lower HTF outlet temperature results in a lower radiated heat flux, meaning more TPV area is required to convert the thermal power to electrical power. Therefore, there is a tradeoff between higher $FOM_{P,discharge}$ and higher cost of installing additional TPV cells.

Figure 5(e) and (f) demonstrate this tradeoff. Using the same storage duration of $\tau = 31.6$ hours, as we increase the flowrate factor, the $FOM_{P,discharge}$ increases. Particularly, we note that in the base case (flowrate factor $f = 1$) we get an $FOM_{P,discharge}$ of 0.4. Increasing the flowrate factor to 3 (meaning the flowrate can increase to 3x the nominal flowrate) allows $FOM_{P,discharge}$ to reach 0.9 (equivalently, a constant power duration of 28.4 hours out of 31.6 hours). However, this requires approximately 40% more TPV cells (since the outlet temperature decreases with time), increasing the capital expenditure of the system.

We have thus far only considered a storage duration of 31.6 hours, so next we expand our analysis to durations ranging from 10 to 100 hours, as shown in Figure 5(g) and (h). As seen, the systems with longer storage duration times require a lower flowrate factor $f$ to reach an FOM of 0.9, while systems with shorter duration require higher $f$ to reach an FOM of 0.9. This can be explained by the different outlet temperature profiles for different storage durations. The longer storage durations already have a more uniform discharge profile without any modifications (at $f = 1$) so only minimal modifications are necessary. In contrast, shorter storage durations have a faster decline in outlet temperature as their nominal flowrate is higher, which makes achieving a constant power output more difficult.

Next, we note that the percent increase in TPV area added (relative to $f = 1$ for each storage duration) is independent of the storage duration. To understand this better, we consider a flowrate factor $f$ – we can calculate the threshold $\Delta T_{f,thres} = \frac{P}{c_{p,Sn}\dot{m}_f(t=0)f} = \frac{c_{P,Sn}\dot{m}_f(t=0)500°C}{c_{P,Sn}\dot{m}_f(t=0)f} = \frac{500°C}{f}$. Because the TPV area is calculated based on $\Delta T_{f,thres}$, the additional TPV area required is independent of the storage duration and only depends on $f$.

In summary, in this sub-section we introduced a technique for keeping discharge electrical power constant. First, we keep output thermal power constant by counterbalancing the dropping HTF outlet temperature with an increasing HTF mass flowrate. To convert this thermal power to electrical power, we throttle the heat engine at the beginning of discharge to allow for high power output later, when the HTF temperature is low and flowrate high. There is a tradeoff between performance and cost, which should be evaluated with technoeconomic analysis in a future work.

### 4.2.2. Accelerating charging

We can apply the same approach to the second objective of fast charging. For this objective the uniformity of power over time is not as important, rather we care about charging the system as fully as possible within the charging duration. We therefore define a charging FOM as the ratio between the electrical energy input over the charging duration to the energy capacity:

$$FOM_{P,charge} = \frac{\int_0^\tau \dot{m}(t) c_{p,Sn} \Delta T(t)\, dt}{\eta_{charge} E}$$

which is essentially the state of charge of the system after charging is complete. Since $\eta_{charge}$ depends on the type of heating infrastructure used and is generally fixed, for this analysis we will focus on the thermal power in the numerator. Maximizing $FOM_{P,charge}$ therefore corresponds to maximizing the amount of thermal power input into the system.

To motivate the above methodology for charging, if we keep the flowrate constant then as the outlet temperature of HTF from the solid storage increases (as the solid heats up), the amount of thermal power we can supply to the HTF through the heaters drops, since the $\Delta T$ between the HTF outlet temperature and the maximum operating temperature of the system decreases. Therefore, we can increase the flowrate to keep the thermal power input constant.

We evaluate the effectiveness of the methodology presented in the context of the TEGS system. For this analysis, we consider a graphite block of the same dimensions as the previous subsection (length $L = 10m$, diameter $D_C = 0.2m$, tin channel diameter $D_{Sn} = 0.02m$, determined from dimensional analysis). For the accelerated charging analysis, we first consider a short charging duration of $\tau = 5$ hours, which corresponds to, for example, the amount of time abundant solar might be available during the day. The nominal flowrate for such a system would be $\dot{m}_{nom} = \frac{E/5h}{c_{p,Sn}(500°\text{C})}$. To accelerate charging, we define a similar maximum flowrate $\dot{m}_{max}$ and a flowrate factor $f = \frac{\dot{m}_{max}}{\dot{m}_{nom}}$. The flowrate is allowed to increase with time as the outlet temperature drops, to keep the power input at its maximum (rated) value, until reaching this maximum flowrate. The results for the given system, with 5 hours of charging duration and various flowrate factors, are shown in Figure 6(a)-(d). Figure 6(e) shows how $FOM_{P,charge}$ varies for other charging durations and flowrate factors.

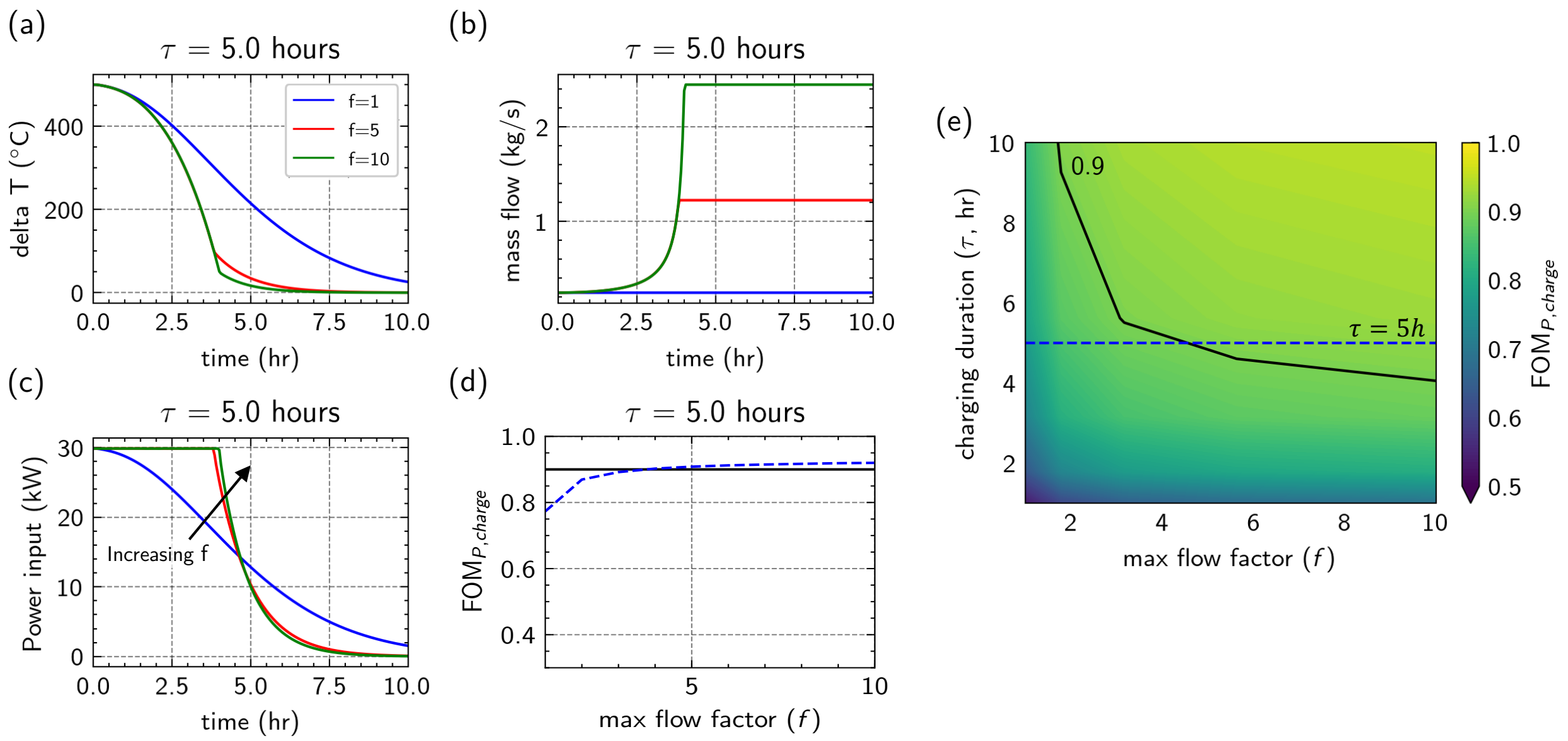

*Figure 6: Demonstration of how flow ramping counteracts the lower tin $\Delta T$ and ensures charging at the maximum rated charging power. First considering a charge duration of 5 hours, (a) $\Delta T$ of the tin outlet vs. inlet to the storage medium, (b) flowrate of tin as a function of time, (c) thermal power of the tin output from the storage medium. (d) The impact of allowing the flowrate to increase to a certain factor on the charging FOM, for $\tau = 5h$. (e) Charge FOM for various storage durations and flowrate factors, with solid line an isocontour of $FOM_{P,charge}$=0.9, and the dashed line showing where $\tau = 5$ lies.*

As seen in Figure 6(d), increasing the flowrate factor from 1 to 5 improves $FOM_{P,charge}$ from 75% to 90%, with nominal improvements above 5x. The specific response of the thermal storage system to various flowrate factors is shown in Figure 6(a)-(c). When considering various charging durations from 1 to 10 hours (Figure 6(e)), the accelerated charging is easier at longer durations since their charging FOMs are already high with no modifications (at $f = 1$), similar to the phenomenon observed in discharge power uniformity. We note that for all durations >4 hours, we can nearly fully charge the system (>90%) with flowrate multipliers of 10 or lower. For shorter durations, full charge could be accomplished with higher flowrate factors, but other techniques such as changing the geometry of the channels might be required instead.

In summary, accelerating charging by increasing HTF flowrate to establish a near-constant power input is promising, as it has the potential to reach almost complete charging in a short duration. Although it increases pressure drop, it requires minimal infrastructure to implement (only additional pumping throughput), so there is little tradeoff between performance and cost. Other options for accelerating charging are presented in Section S7, but are shown to be not effective for various reasons.

In our analysis thus far, we have only considered a single channel-embedded cylinder. In the following section, we will analyze a larger block consisting of multiple channels and will add modeling of radiation between blocks, to determine how different configurations impact performance, as well as how to improve poorly performing configurations.

## 4.3. *Large-scale modeling of a thermal storage system*

To analyze a large array of blocks, approximations must be made to reduce the computational expense of solving the full fluid and temperature profiles within each block. We employ a porous media approximation for each block. The details of the porous media approximation are provided in Section 5.2, with the essence being that the porous block is made to exhibit equivalent heat transfer and fluid flow as the channel-embedded block by matching porosity, permeability, and the solid-fluid heat transfer coefficient. The resulting approximation shows good agreement to the fully coupled channel-embedded model with 2-3 orders of magnitude faster computation (see Section S6).

To show how this porous media approximation can be used for design, we consider the TEGS system with graphite as the storage medium and tin as the HTF. We consider large rectangular graphite blocks of dimensions 1m x 1m x 4m, which was determined by quotes from manufacturers and structural limitations of resting the blocks on porous insulating materials. From the previous dimensional analysis, we found that 0.02m tin channel diameter with 0.2m spacing is ideal – therefore each graphite block must have a 5x5 array of tin channels to allow a spacing of 0.2m (as shown in Figure 1(c)). We also noticed that the ideal tin flow path length was at least 10m, but since each block is only 4m high, blocks may have to be connected in series. Due to the high temperatures of the storage system (~2000°C), the blocks also emit significant radiation which must be included in the simulation. Radiation can detract from (dis)charge performance by helping make the system more isothermal, reducing the outlet temperature of the tin.

We consider a grid-scale system of ~200 MWh of thermal energy storage capacity, corresponding to 100 blocks. Using the porous media approximation, we investigate 5 different block arrangement configurations, as shown in Figure 7.

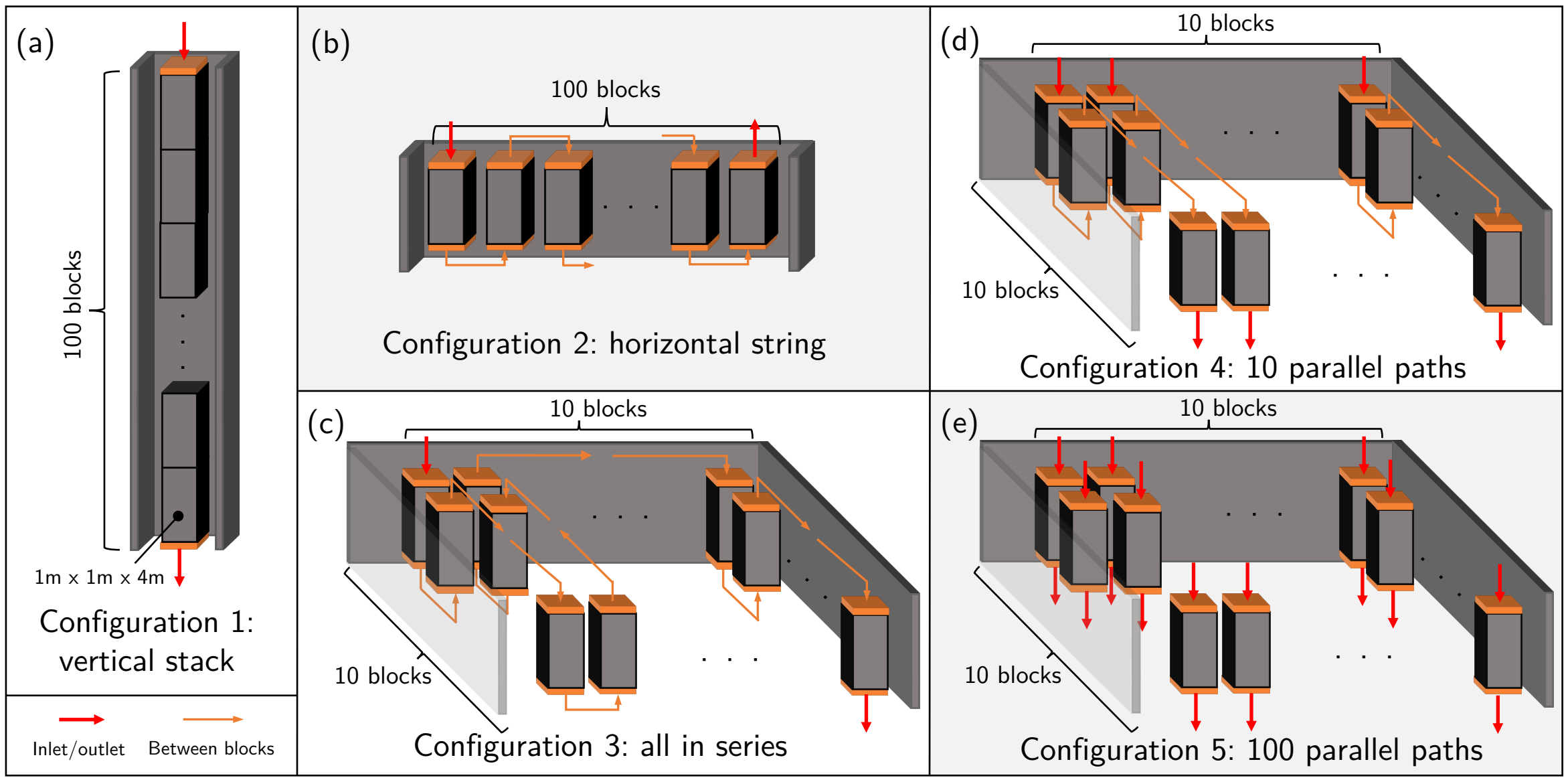

*Figure 7: Comparison of various block configurations for the large-scale grid storage simulation. 100 blocks of 1m x 1m x 4m are considered here. Insulation surrounds the blocks and is occasionally shown as transparent for clarity. (a) Shows all 100 blocks stacked vertically, (b) shows 100 blocks placed in a line, (c) shows a 10 x 10 grid of blocks with a single flow path connecting all of them, (d) shows a 10 x 10 grid of blocks with 10 flow paths, and (e) shows a 10 x 10 grid of blocks with 100 independent flow paths.*

The first configuration (Figure 7(a)) includes all blocks stacked vertically, which ensures blocks are thermally isolated from each other since each block radiates only to the surrounding insulation, but is impractical due to the immense height of the resulting system. The second configuration (Figure 7(b)) has all blocks arranged in a string, which ensures each block can only radiate to 2 adjacent blocks, limiting interaction. Configurations 1 and 2 require a significant amount of insulation due to their long aspect ratio, but they are optimal from a heat transfer standpoint.

Next, we consider 3 more practical configurations using a grid of blocks. Configuration 3 (Figure 7(c)) arranges a 9x11 grid with all blocks in series such that the tin inlet is only on the top left block and the tin outlet is only on the bottom right (9x11 was used to ensure the inlet and outlet are maximally separated). Configuration 4 (Figure 7(d)) consists of a 10x10 grid with 10 parallel paths for tin flow, such that tin enters in the top row and exits in the bottom row. Configuration 5 (Figure 7(e)) features a 10x10 grid with all blocks in parallel, such that tin enters and exits from each block. Note that configurations 3, 4, and 5 have significantly less outer surface area than 1 and 2, reducing the amount of insulation required.

Now, we consider discharging performance over a duration of 20 hours. The results for the tin outlet temperature as a function of time for the 5 different configurations are shown in Figure 8(a). As seen, configurations 1 and 2 have very similar performance. While no blocks can see each other in configuration 1, in configuration 2, 2 sides of each block are exposed to another block, but because each block is not significantly different in temperature from adjacent blocks, this has limited influence on temperature, so its performance is similar to configuration 1.

Configurations 3 and 4 are also similar. In both cases, most blocks are exposed on all 4 sides to other blocks. The temperature difference between adjacent blocks is possibly much higher than in configuration 2, which helps explain the difference in performance. GIFs of the temperature evolution of the blocks for configurations 3 and 4 are present in Section S8. Configuration 5 is worse than configurations 3 and 4. This difference can be partially explained by the short tin flow path of 4m in configuration 5, which is less than the 10m ideal discovered through dimensional analysis.

As seen, the discharge performance of the 10x10 grids needs to be improved. We thus employ the technique outlined in Section 4.2.1 of increasing flowrate and TPV area with time to improve the power uniformity during discharge. As seen in Figure 8(b), this methodology vastly improves performance and even makes the 10x10 case outperform the vertically stacked case. Notably, only a small increase in maximum flowrate (1 to 1.25) enables the configuration to outperform the vertically stacked case in terms of duration of constant discharge power, with only a minimal cost increase (additional pumping costs and ~10% more TPV area). With a maximum flowrate of 5x the nominal (requiring ~50% more TPV area), we achieve constant power output for ~90% of the 20-hour discharge cycle (18 hours), as seen in Figure 8(c).

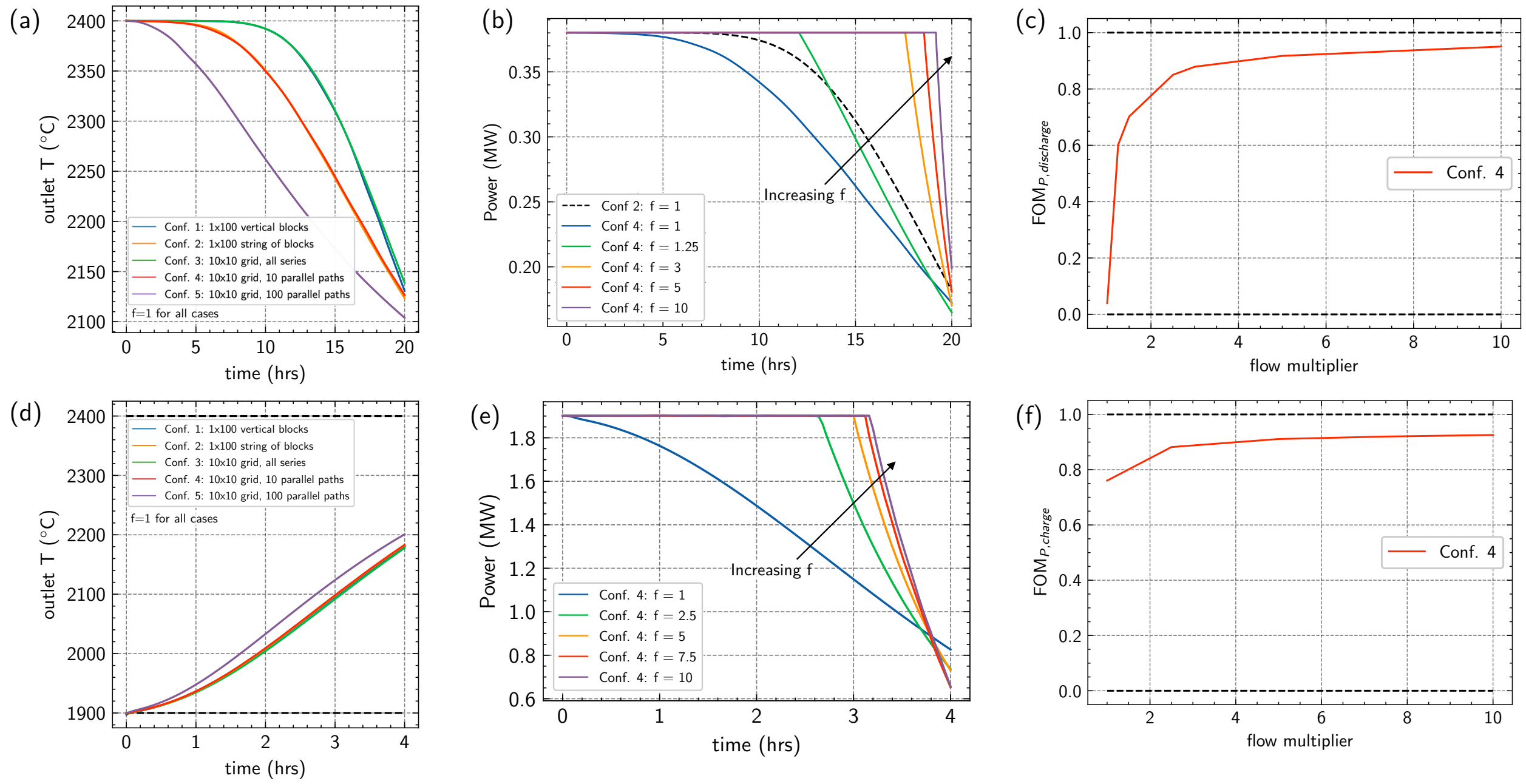

*Figure 8: Multi-block simulations of 100 blocks (200MWh of thermal energy capacity), considering 20 hours of discharge and 4 hours of charging. (a) Comparison of the outlet temperature of tin vs. time for the 5 block configurations described previously. The blue and green lines nearly overlap, as do the yellow and red lines. (b) The outlet power from the tin to the discharging infrastructure, for the 1x100 reference case (conf 2), as well as the 10x10 grid case (conf 4) implementing the flowrate ramping technique presented previously (for f = 1 through 10), showing the 10x10 grid can perform better than the 1x100 case when implementing flowrate ramping. (c) Demonstrates the impact of increasing maximum flowrate on the discharging FOM for the 10x10 grid. GIFs of temperature profile vs. time for configuration (3) and (4) during discharge are shown in Figure S5. (d) Comparison of the outlet temperature of tin vs. time for the 5 block configurations during charging. (e) The inlet power from the charging infrastructure to the tin, for various flowrate multipliers in configuration 4. (f) The impact of increasing flowrate ramping factor on the charging FOM for the 10x10 grid.*

Next, we consider charging performance over a duration of 4 hours. The tin outlet temperature as a function of time for the 5 different configurations are shown in Figure 8(d). As seen, the difference between the configurations is not as drastic as in discharging, suggesting that the faster rate of charging causes adjacent blocks to reach similar temperatures quicker, limiting the effect of radiation between blocks. However, the 10x10 grid with all blocks in parallel (configuration 5) still performs worse than all other cases, again demonstrating the importance of the dimensional analysis conducted previously showing the minimum tin path length of 10m.

Further, it is evident that charging performance needs to be improved, since the tin outlet temperature has not reached 2400C within the 4 hours (which would indicate a fully charged system). For accelerating charging, we use the technique introduced in Section 4.2.2 of increasing flowrate above nominal as the temperature rises to create a constant power input. Figure 8(e) shows how increasing the flowrate multiplier f allows for longer durations of constant input power, accelerating the charging speed. Figure 8(f) shows that the block is nearly completely charged within 4 hours by allowing the flowrate to increase five-fold above nominal, which improves charging performance while only increasing pumping costs.

Overall, these simulations show that for a grid of blocks, the choice of flow path is minimally important as long as the tin flow path length is higher than the minimum determined by dimensional analysis. If no modifications to flowrate were used, a thermally isolated string of blocks would be ideal, but this would be impractical due to the drastically high insulation costs. Although the grid is worse due to radiation between the blocks, the techniques developed in this work can be utilized to enhance performance.

# 5. Experimental procedures

## *5.1. Numerical modeling of a single channel-embedded cylinder*

COMSOL Multiphysics® was used for coupled fluid flow and heat transfer modeling. Low-quality graphite was used was the storage medium while liquid tin was used as the heat transfer fluid. The relevant material properties of graphite and tin are present in the table below. If ranges were provided, a temperature-dependent property was used in the model varying from 1900 to 2400C.

| Material | Property | Value |
|---|---|---|
| Low-quality graphite | Thermal conductivity ($k$) | 10 W/m/K |
| | Specific heat ($c_p$) | 2000 J/kg/K |
| | Density ($\rho$) | 1700 kg/m$^3$ |
| Liquid tin | Thermal conductivity ($k$) | 60-65 W/m/K |
| | Specific heat ($c_p$) | 247-250 J/kg/K |
| | Density ($\rho$) | 4800 kg/m$^3$ |
| | Dynamic viscosity ($\mu$) | 0.001 Pa·s |

An axially symmetric 2D model was used with the outer boundary being adiabatic. Other dimensions and flowrates were determined based on user inputs. As mentioned, laminar flow was used as the primary purpose was to accurately model heat transfer, and turbulent flow only minimally impacts heat transfer as detailed in Section S4.

For the example case presented in Section 3, a block of diameter 0.2m, length 10m, and channel diameter 0.02m was used, with a tin flowrate of 0.04 kg/s, and other material properties given above.

During discharging we have assumed a constant $\eta_{discharge}$ of 0.4, and during charging we have assumed a constant $\eta_{charge}$ of 0.99.

## *5.2. Porous media approximation*

Figure 9 shows how a block of 25 channels can be approximated as a single block with porosity.

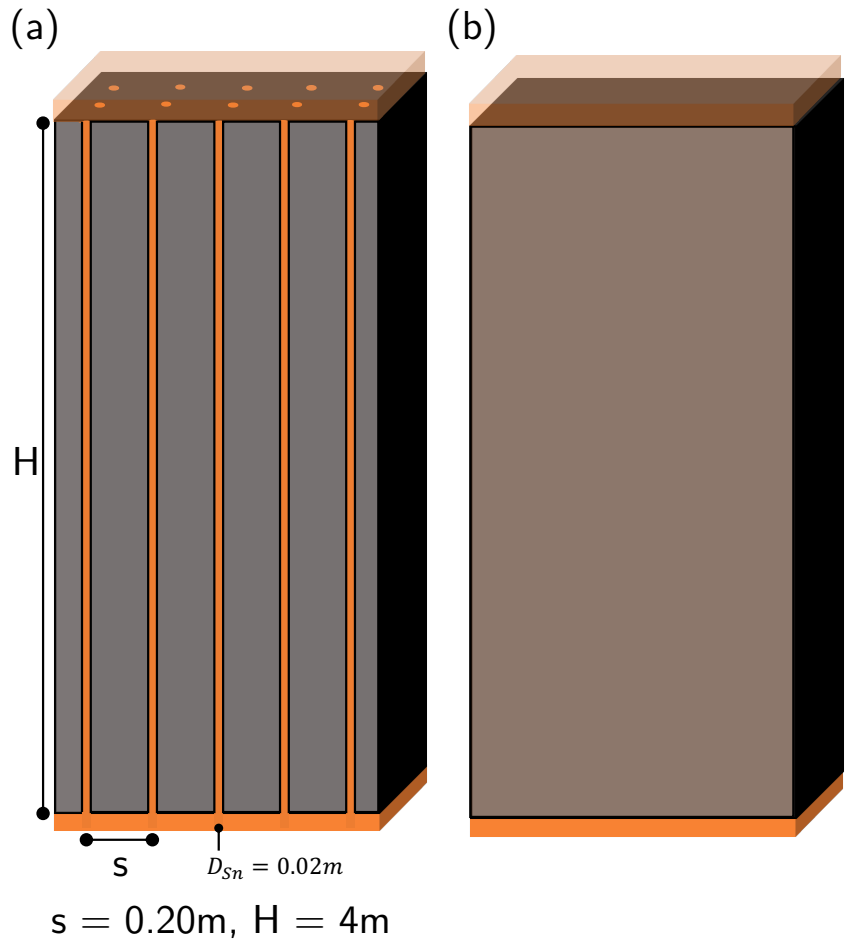

*Figure 9: Schematic of porous media approximation showing (a) Cross section of 1m x 1m x 4m graphite block with 25 tin channels embedded of diameter 2cm, and (b) porous media approximation of the same block, where two temperature fields are tracked in the entire block, with parameters matched to achieve equivalent fluid and thermal response.*

For a porous block to exhibit equivalent heat transfer and fluid flow as a channel-embedded block, we must match some parameters.[28] First is the porosity, such that the volume fraction of liquid and solid is the same. Second is the permeability, which is tuned to match the pressure drop in the reference block. Namely, the porous media fluid flow adds an additional source term in the momentum equation:

$$S = \kappa^{-1}\mu + \left(\frac{1}{2}\frac{\rho|u|}{\sqrt{\kappa}}\right)u$$

where $\kappa$ is the permeability, $\mu$ is the viscosity, and $u$ is the velocity. The permeability was tuned such that the pressure drops matched.

Next, the porous media approximation keeps track of two temperature fields, one for fluid and one for solid. They are coupled with a heat source/sink term involving a relative heat transfer coefficient. The governing heat transfer equations in the solid and fluid, respectively, are:

$$(1-\epsilon_p)\rho_s c_{p,s}\frac{\partial T_s}{\partial t} + \left(-(1-\epsilon_p)k_s\nabla^2 T_s\right) = h_{sf}\left(\frac{A}{V}\right)\left(T_f - T_s\right)$$

$$\epsilon_p\rho_f c_{p,f}\frac{\partial T_f}{\partial t} + \epsilon_p\rho_f c_{p,f}(u_f\cdot\nabla)T_f + (-\epsilon_p k_f\nabla^2 T_f) = h_{sf}\left(\frac{A}{V}\right)(T_s - T_f)$$

where $\epsilon_p$ is the porosity, $s$ subscripts indicate solid while $f$ indicates fluid. $A$ and $V$ are the pore surface area and volume, which are calculated directly from the geometry of the reference block. $h_{sf}$ is the heat transfer coefficient between the solid and fluid. In our case, it includes convection and conduction resistances. Namely,

$$h_{sf} = \frac{1}{(R_{conv} + C_1 R_{cond})A}$$

$$R_{conv} = \frac{1}{hA} = \frac{1}{Nu\frac{k}{D}A}, \;\; R_{cond} = \frac{ln\left(\frac{r_o}{r_i}\right)}{2\pi kL}$$

Where $C_1$ is a correction factor to both convert the cylindrical conduction resistance to rectangular and to convert the conduction resistance to a convection resistance, as was done in Xu et al.[29] Note that because the lateral conduction is included in the heat transfer coefficient $h_{sf}$ through the $R_{cond}$ term, the thermal conductivity $k_s$ in the governing heat transfer equation is anisotropic, with 0 laterally and the nominal values axially, to avoid double-counting the conduction contributions. Once the above values are obtained, a coarse-grained porous media approximation may be used.

To validate this approximation, we consider the TEGS system, and model a large graphite block with 25 tin channels. We compare the tin outlet temperature vs. time for both the full geometry case and the porous media approximation for a variety of flowrates. This comparison is shown in Figure S6. As seen, the porous media approximation offers good agreement with 2-3 orders of magnitude faster computation.

### *5.3. Simulation parameters for large-scale modeling*

For each of the configurations, we consider surface-to-surface radiation between blocks and the insulation, which is externally adiabatic. We use a spacing of 10cm between the blocks and insulation, and 20cm between blocks. We assume all surfaces have emissivity 0.9, which is slightly higher than the reported emissivity of graphite at high temperatures,[30,31] so overestimates the heat transfer between blocks. Where possible, the flow configuration is defined such that if the outlet of one block is at the bottom (top), the inlet of the subsequent block is also at the bottom (top). This reflects the real-world flow conditions. Only a 10x10 grid is considered in this work as this is the most optimal grid to minimize outer surface area for insulation. Other grids could be considered by repeating the analysis and taking advantage of the faster simulation time of the porous medium approximation.

## 6. Conclusions and future work

In summary, this work introduces design principles for sensible heat solid thermal storage systems to improve their (dis)charge performance. We consider the channel-embedded configuration, which features channels of heat transfer fluid embedded within the solid storage medium. Typically, these systems may have poor performance, facing challenges in output power uniformity during discharge and slow charging rates. Further, if the sensible heating occurs at high temperatures as required for high round-trip efficiency, radiation between blocks can degrade performance. This suggests design optimization is necessary. However, grid-scale versions of this system are computationally intractable to model fully. Therefore, in this work we have utilized a hierarchical approach to design optimization, developing design principles for channel-embedded thermal storage systems at each step of the hierarchy. We then apply these design principles to a specific system that uses graphite as the storage medium and liquid tin as the heat transfer fluid (called thermal energy grid storage or TEGS).

First, we investigate the impact of geometric parameters on performance using dimensional analysis. We define an FOM based on temperature uniformity with time and derive the non-dimensional groups that impact this FOM for a general system. To demonstrate how this analysis can be used for a specific system, we consider the TEGS system, which allows us to simplify some terms and develop a relationship between storage medium diameter, heat transfer fluid flow path length, and FOM. Specifically, for a storage medium of thermal conductivity 10 W/m/K, we find a diameter of 20cm and length of 10m achieves a temperature FOM of 90% for a discharge duration of 30 hours. While we have generated these numbers for TEGS specifically, the dimensional analysis is generalizable to other channel-embedded thermal energy storage systems using different materials and storage durations.

Next, we investigate how varying the operation of the storage system can help optimize discharging and charging performance. For discharge, we define an FOM based on the time the storage system can provide the nominal power compared to the rated storage duration. We find that increasing HTF flowrate as the outlet temperature drops enables near-constant power output, but requires throttling of the heat engine. Using the TEGS system as an example, we find that allowing flowrate to increase by 5x and TPV area to increase 50% allows >90% discharge FOM. For charging, we define an FOM based on the fraction of storage capacity able to be charged in a certain charging duration, i.e. the state of charge after the charging duration. We find that we can apply the same methodology from discharging of ramping up flowrate as temperature difference drops, to keep input power constant at its maximum value. For TEGS, allowing flowrate to ramp up by 5x allows 90% charging within 5 hours.

Finally, we model the full grid-scale system and propose a porous media approximation that is calibrated to match the heat transfer and fluid flow properties of the fully coupled case to accelerate computation. We apply this to TEGS to find and consider 100 1m x 1m x 4m commercially available graphite blocks (corresponding to a thermal energy capacity of 200MWh) arranged in various configurations to determine the effect of radiation between the blocks. We first find that blocks should be connected in series to achieve the minimum tin flow path as determined by dimensional analysis (at least 10m). We then find that the 10x10 grid arrangement of blocks (minimizing insulation costs) underperforms compared to arrangements featuring blocks that are thermally isolated from each other. By utilizing the insights generated in previous parts, this thermal performance can be improved to near its maximum, and we demonstrate discharging and charging FOMs of >90%.

In this work we have primarily studied heat transfer considerations and have briefly discussed their impact on cost where appropriate. A more rigorous technoeconomic analysis evaluating each proposed improvement to performance is warranted and should be the focus of a future study. Further, while we have considered discharging and charging in isolation, it is also important to consider the transition between the two. While a detailed analysis is outside the scope of this work, many results in the literature suggest that flow should be reversed between discharge and charge, to ensure thermoclines developed in the storage media are maintained.[32–34] Longer-duration studies where the system is charged and discharge multiple times should be the subject of a future work.

Further, it would be useful to model the grid integration of such a thermal energy storage system with realistic power input/output profiles. This would involve further modeling and approximations to the large-scale model to enable longer duration simulations.

Overall, this work develops design principles to improve the performance of cheap thermal energy grid storage systems, which in turn facilitates its adoption and helps increase renewables penetration for a cleaner future.

## 7. Data availability

Data and code are available on GitHub.[35]

## 8. Acknowledgements

This material is based upon work supported by the National Science Foundation Graduate Research Fellowship under Award Nos. 1745302 and 2141064.

## 9. Author contributions

Conceptualization, S.V., C.K., A.H.; Methodology, S.V., C.K., K.B., A.L., A.S., A.H.; Investigation, S.V.; Writing – Original Draft, S.V.; Writing – Review and Editing, S.V., K.B., A.L., A.S., A.H.; Supervision, A.H.

## 10. Declaration of interests

A.H. has founded a company for thermal energy storage based on the TEGS system discussed in this work, and therefore has a financial interest.

# 11. References

(1) Dowling, J. A.; Rinaldi, K. Z.; Ruggles, T. H.; Davis, S. J.; Yuan, M.; Tong, F.; Lewis, N. S.; Caldeira, K. Role of Long-Duration Energy Storage in Variable Renewable Electricity Systems. *Joule* **2020**, *4* (9), 1907–1928. https://doi.org/10.1016/j.joule.2020.07.007.

(2) Ziegler, M. S.; Mueller, J. M.; Pereira, G. D.; Song, J.; Ferrara, M.; Chiang, Y.-M.; Trancik, J. E. Storage Requirements and Costs of Shaping Renewable Energy Toward Grid Decarbonization. *Joule* **2019**, *3* (9), 2134–2153. https://doi.org/10.1016/j.joule.2019.06.012.

(3) Glatzmaier, G. *Developing a Cost Model and Methodology to Estimate Capital Costs for Thermal Energy Storage*; NREL/TP-5500-53066; National Renewable Energy Lab. (NREL), Golden, CO (United States), 2011. https://doi.org/10.2172/1031953.

(4) Laughlin, R. B. Pumped Thermal Grid Storage with Heat Exchange. *J. Renew. Sustain. Energy* **2017**, *9* (4), 044103. https://doi.org/10.1063/1.4994054.

(5) Armstrong, R.; Chiang, Y.-M.; Gruenspecht, H. *The Future of Energy Storage: Chapter 4 - Thermal Energy Storage*; MIT Future of; Massachusetts Institute of Technology, 2022; pp 113–146. https://energy.mit.edu/wp-content/uploads/2022/05/The-Future-of-Energy-Storage.pdf.

(6) Bauer, T. Chapter 1 - Fundamentals of High-Temperature Thermal Energy Storage, Transfer, and Conversion. In *Ultra-High Temperature Thermal Energy Storage, Transfer and Conversion*; Datas, A., Ed.; Woodhead Publishing Series in Energy; Woodhead Publishing, 2021; pp 1–34. https://doi.org/10.1016/B978-0-12-819955-8.00001-6.

(7) Amy, C.; Seyf, H. R.; Steiner, M. A.; Friedman, D. J.; Henry, A. Thermal Energy Grid Storage Using Multi-Junction Photovoltaics. *Energy Environ. Sci.* **2019**, *12* (1), 334–343. https://doi.org/10.1039/C8EE02341G.

(8) Herrmann, U.; Kelly, B.; Price, H. Two-Tank Molten Salt Storage for Parabolic Trough Solar Power Plants. *Energy* **2004**, *29* (5), 883–893. https://doi.org/10.1016/S0360-5442(03)00193-2.

(9) Alva, G.; Lin, Y.; Fang, G. An Overview of Thermal Energy Storage Systems. *Energy* **2018**, *144*, 341–378. https://doi.org/10.1016/j.energy.2017.12.037.

(10) Xie, B.; Baudin, N.; Soto, J.; Fan, Y.; Luo, L. Chapter 10 - Thermocline Packed Bed Thermal Energy Storage System: A Review. In *Renewable Energy Production and Distribution*; Jeguirim, M., Ed.; Advances in Renewable Energy Technologies; Academic Press, 2022; Vol. 1, pp 325–385. https://doi.org/10.1016/B978-0-323-91892-3.24001-6.

(11) ELSihy, Els. S.; Liao, Z.; Xu, C.; Du, X. Dynamic Characteristics of Solid Packed-Bed Thermocline Tank Using Molten-Salt as a Heat Transfer Fluid. *Int. J. Heat Mass Transf.* **2021**, *165*, 120677. https://doi.org/10.1016/j.ijheatmasstransfer.2020.120677.

(12) Rodrigues, F. A.; de Lemos, M. J. S. Discharge Effectiveness of Thermal Energy Storage Systems. *Appl. Therm. Eng.* **2022**, *209*, 118232. https://doi.org/10.1016/j.applthermaleng.2022.118232.

(13) Rodrigues, F. A.; de Lemos, M. J. S. Effect of Porous Material Properties on Thermal Efficiencies of a Thermocline Storage Tank. *Appl. Therm. Eng.* **2020**, *173*, 115194. https://doi.org/10.1016/j.applthermaleng.2020.115194.

(14) Marti, J.; Geissbühler, L.; Becattini, V.; Haselbacher, A.; Steinfeld, A. Constrained Multi-Objective Optimization of Thermocline Packed-Bed Thermal-Energy Storage. *Appl. Energy* **2018**, *216*, 694–708. https://doi.org/10.1016/j.apenergy.2017.12.072.

(15) Esence, T.; Bruch, A.; Molina, S.; Stutz, B.; Fourmigué, J.-F. A Review on Experience Feedback and Numerical Modeling of Packed-Bed Thermal Energy Storage Systems. *Sol. Energy* **2017**, *153*, 628–654. https://doi.org/10.1016/j.solener.2017.03.032.

(16) Wu, M.; Li, M.; Xu, C.; He, Y.; Tao, W. The Impact of Concrete Structure on the Thermal Performance of the Dual-Media Thermocline Thermal Storage Tank Using Concrete as the Solid Medium. *Appl. Energy* **2014**, *113*, 1363–1371. https://doi.org/10.1016/j.apenergy.2013.08.044.

(17) Jian, Y.; Falcoz, Q.; Neveu, P.; Bai, F.; Wang, Y.; Wang, Z. Design and Optimization of Solid Thermal Energy Storage Modules for Solar Thermal Power Plant Applications. *Appl. Energy* **2015**, *139*, 30–42. https://doi.org/10.1016/j.apenergy.2014.11.019.

(18) Jian, Y.; Bai, F.; Falcoz, Q.; Xu, C.; Wang, Y.; Wang, Z. Thermal Analysis and Design of Solid Energy Storage Systems Using a Modified Lumped Capacitance Method. *Appl. Therm. Eng.* **2015**, *75*, 213–223. https://doi.org/10.1016/j.applthermaleng.2014.10.010.

(19) Doretti, L.; Martelletto, F.; Mancin, S. A Simplified Analytical Approach for Concrete Sensible Thermal Energy Storages Simulation. *J. Energy Storage* **2019**, *22*, 68–79. https://doi.org/10.1016/j.est.2019.01.029.

(20) Prasad, L.; Niyas, H.; Muthukumar, P. Performance Analysis of High Temperature Sensible Heat Storage System during Charging and Discharging Cycles; ndian Institute of Technology Bombay, Mumb, 2013; pp 1240–1247.

(21) Doretti, L.; Martelletto, F.; Mancin, S. Numerical Analyses of Concrete Thermal Energy Storage Systems: Effect of the Modules' Arrangement. *Energy Rep.* **2020**, *6*, 199–214. https://doi.org/10.1016/j.egyr.2020.07.002.

(22) Kuang, R.; Huang, N.; Chen, G.; Tan, J.; Liu, J.; Shen, Y. Numerical Analysis of Discharging Stability of Basalt Fiber Bundle Thermal Energy Storage Tank. *Energy Rep.* **2022**, *8*, 13014–13022. https://doi.org/10.1016/j.egyr.2022.09.115.

(23) LaPotin, A.; Schulte, K. L.; Steiner, M. A.; Buznitsky, K.; Kelsall, C. C.; Friedman, D. J.; Tervo, E. J.; France, R. M.; Young, M. R.; Rohskopf, A.; Verma, S.; Wang, E. N.; Henry, A. Thermophotovoltaic Efficiency of 40%. *Nature* **2022**, *604* (7905), 287–291. https://doi.org/10.1038/s41586-022-04473-y.

(24) Verma, S.; Adams, M.; Foxen, M.; Sperry, B.; Yee, S.; Henry, A. High-Temperature Thermal Conductivity Measurements of Macro-Porous Graphite. arXiv February 2, 2023. https://doi.org/10.48550/arXiv.2301.03440.

(25) Kelsall, C. C.; Buznitsky, K.; Henry, A. Technoeconomic Analysis of Thermal Energy Grid Storage Using Graphite and Tin. *ArXiv210607624 Phys.* **2021**.

(26) Sepulveda, N. A.; Jenkins, J. D.; Edington, A.; Mallapragada, D. S.; Lester, R. K. The Design Space for Long-Duration Energy Storage in Decarbonized Power Systems. *Nat. Energy* **2021**, *6* (5), 506–516. https://doi.org/10.1038/s41560-021-00796-8.

(27) Eikeland, O. F. Enhancing Decision-Making in the Electric Power Sector with Machine Learning and Optimization. Doctoral thesis, UiT Norges arktiske universitet, 2023. https://munin.uit.no/handle/10037/31514 (accessed 2024-01-31).

(28) Zhu, Q.; Pishahang, M.; Caccia, M.; Kelsall, C. C.; LaPotin, A.; Sandhage, K. H.; Henry, A. Validation of the Porous Medium Approximation for Hydrodynamics Analysis in Compact Heat Exchangers. *J. Fluids Eng.* **2022**, *144* (8). https://doi.org/10.1115/1.4053898.

(29) Xu, B.; Li, P.-W.; Chan, C. L. Extending the Validity of Lumped Capacitance Method for Large Biot Number in Thermal Storage Application. *Sol. Energy* **2012**, *86* (6), 1709–1724. https://doi.org/10.1016/j.solener.2012.03.016.
(30) Shu, S.; Wu, T.; Wang, Z.; Zhang, Y.; Yang, Z.; Liang, H. Research on the Normal Emissivity of Graphite between 150 and 500 °C by an Infrared Camera for Nuclear Fusion Devices. *Nucl. Mater. Energy* **2022**, *31*, 101182. https://doi.org/10.1016/j.nme.2022.101182.
(31) Ren, D.; Tan, H.; Xuan, Y.; Han, Y.; Li, Q. Apparatus for Measuring Spectral Emissivity of Solid Materials at Elevated Temperatures. *Int. J. Thermophys.* **2016**, *37* (5), 51. https://doi.org/10.1007/s10765-016-2058-9.
(32) *Thermal Storage System Concentrating Solar-Thermal Power Basics*. Energy.gov. https://www.energy.gov/eere/solar/thermal-storage-system-concentrating-solar-thermal-power-basics (accessed 2024-02-02).
(33) Dickinson, J. S.; Buik, N.; Matthews, M. C.; Snijders, A. Aquifer Thermal Energy Storage: Theoretical and Operational Analysis. *Géotechnique* **2009**, *59* (3), 249–260. https://doi.org/10.1680/geot.2009.59.3.249.
(34) Riahi, S.; Saman, W. Y.; Bruno, F.; Belusko, M.; Tay, N. H. S. Impact of Periodic Flow Reversal of Heat Transfer Fluid on the Melting and Solidification Processes in a Latent Heat Shell and Tube Storage System. *Appl. Energy* **2017**, *191*, 276–286. https://doi.org/10.1016/j.apenergy.2017.01.091.
(35) *shomikverma/graphite-discharge: code for graphite discharge modeling project*. GitHub. https://github.com/shomikverma/graphite-discharge (accessed 2023-07-26).
(36) *CRC Handbook of Chemistry and Physics, 100th Edition*, 100th edition.; Rumble, J., Ed.; CRC Press: Boca Raton London New York, 2019.
(37) Informatics, N. O. of D. and. *NIST Chemistry WebBook*. https://webbook.nist.gov/chemistry/ (accessed 2024-02-10).
(38) Taler, D. Heat Transfer in Turbulent Tube Flow of Liquid Metals. *Procedia Eng.* **2016**, *157*, 148–157. https://doi.org/10.1016/j.proeng.2016.08.350.
(39) Brockmann, H. Analytic Angle Factors for the Radiant Interchange among the Surface Elements of Two Concentric Cylinders. *Int. J. Heat Mass Transf.* **1994**, *37* (7), 1095–1100. https://doi.org/10.1016/0017-9310(94)90195-3.

# Supplementary Information for "Designing for effective heat transfer in a solid thermal energy storage system"

Shomik Verma, Colin Kelsall, Kyle Buznitsky, Alina LaPotin, Ashwin Sandeep, Asegun Henry*

Department of Mechanical Engineering, Massachusetts Institute of Technology, 77 Massachusetts Avenue, Cambridge, MA 02139, U.S.A.
* ase@mit.edu

## Table of Contents

## S1. Thermal energy storage materials data

1 https://www.statista.com/statistics/219339/us-prices-of-cement/
2 https://www.forbes.com/home-improvement/foundation/concrete-slab-cost/
3 https://www.statista.com/statistics/219381/sand-and-gravel-prices-in-the-us/#:~:text=In%20the%20United%20States%2C%20the,per%20metric%20ton%20in%202022.
4 https://tradingeconomics.com/commodity/iron-ore
5 https://medium.com/@keruirefractory/exploring-how-much-do-fire-bricks-cost-a93a4530dffd#:~:text=Based%20on%20the%20above%20factors%20and%20other,can%20cost%20between%20$5%20and%20$15%20each.
6 https://www.northerngraphite.com/about-graphite/graphite-pricing/
7 https://www.sciencedirect.com/science/article/abs/pii/S2352152X17302864?via%3Dihub
8 https://bulknaturaloils.com/mineral-oil-usp-90-c1016.html
9 https://www.lme.com/en/metals/non-ferrous/lme-aluminium#Summary
10 https://tradingeconomics.com/commodity/magnesium
11 https://price.metal.com/Silicon
12 https://www.nasdaq.com/market-activity/commodities/hg:cmx
13 https://www.lme.com/en/Metals/Ferrous/LME-Steel-Scrap-CFR-Turkey-Platts
14 https://ycharts.com/indicators/nickel_price#:~:text=Nickel%20Price%20is%20at%20a,42.88%25%20from%20one%20year%20ago.
15 https://ycharts.com/indicators/tin_price
16 https://www.dailymetalprice.com/metalpricecharts.php?c=ga&u=kg&d=240
17 https://www.metal.com/search?keyword=calcium&type=price
18 https://www.metal.com/search?keyword=sodium&type=price
19 https://www.metal.com/search?keyword=potassium&type=price
20 https://www.chemanalyst.com/Pricing-data/quicklime-1505
21 https://businessanalytiq.com/procurementanalytics/index/silica-price-index/

Material data was obtained from the CRC handbook[36] and NIST.[37]

Solid heat capacity was calculated based on solid-phase heat capacity (assuming a constant $c_p$ and a temperature difference of 500°C. Same for liquid phase heat capacity. The phase was determined based on which phase is able to store the most amount of heat, using 500°C for sensible heating, or the latent heat. Many of the natural materials would decompose before melting or shortly after melting, so these were considered to remain solid.

## S2. De-rating the storage duration

We could design a system with a nominal power output $P_{max}$ for $\tau$ hours, but only discharge it for $\tau_{rated}$ hours, where $\tau_{rated}$ is defined as

$$\tau_{rated} = f_{rated}\tau$$

Where $f_{rated} \in (0,1]$. We can quantitatively evaluate this improvement with a figure of merit (FOM) defined as

$$FOM_T = \frac{\int_0^{f_{rated}\tau} \Theta(t)\, dt}{\int_0^{f_{rated}\tau} dt} = \frac{\int_0^{f_{rated}} \Theta^* \ (t^*) dt^*}{\int_0^{f_{rated}} dt^*} \leq 1$$

where $\Theta(t)$ is the non-dimensionalized (dis)charge temperature as a function of time, and $\Theta^*(t^*)$ is the non-dimensionalized (dis)charge temperature as a function of non-dimensionalized time (as shown in Figure 2), such that $\Theta_k^*(t^*) = \Theta_k(t^* \cdot \tau) = \Theta_k(t)$.

We now see some benefits of de-rating – for example, if the storage system in Figure 2 with $k = 5$ was only rated for $f_{rated} = 0.25$ instead of 1, $FOM_T$ would be approximately 1. However, if we compare several cases where $\tau_{rated}$ is constant, as we decrease $f_{rated}$, $\tau$ must increase. In the example above, $\tau$ must be 4 times larger, which means 4 times as many graphite blocks would be required, directly increasing costs. This is shown in Figure S1, where (a) demonstrates the increase in FOM enabled by de-rating (increasing H while keeping U constant), while (b) shows another FOM based on cost (higher H for same U results in higher costs and lower FOM. When we combine these two FOMs, we notice that an $f_{rated}$ value of 1 (dashed diagonal line) maximizes the combined FOM.

The other downside of oversizing the system is that the charging efficacy is lowered. Because the system temperature is always high, the heat transfer into the block during charging is limited due to a smaller $\Delta T$. Therefore the system is unable to provide as much charging power, and the charging duration is lengthened.

We are therefore interested in determining how performance can be maximized based on changes to geometry, operation, and block arrangement in the storage system, while keeping $f_{rated} = 1$.

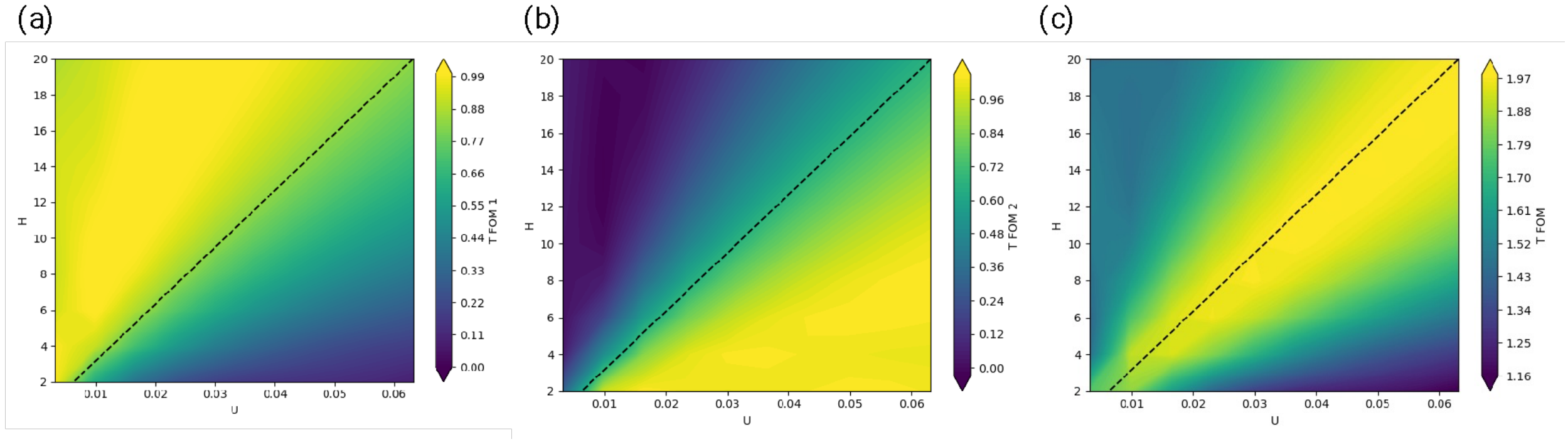

*Figure S1: FOM of derating discharge*

## S3. Dimensional analysis

From the main text, we know that

$$FOM_T = f\left(\frac{L}{D}, \frac{UD}{\alpha_C}, \frac{k_{Sn}}{k_C}\right)$$

We first verify that these are the correct groups, as presented in Figure S2. We assume $\rho_c, c_{p,C}$, and $k_{Sn}$ are constant, so the 3 groups simplify to $\frac{L}{D}$, $\frac{UD}{k_C}$, and $\frac{1}{k_C}$. Changing the values of $L, D, U, k_C$ but keeping the dimensional groups the same resulted in the same tin outlet temperature profile.

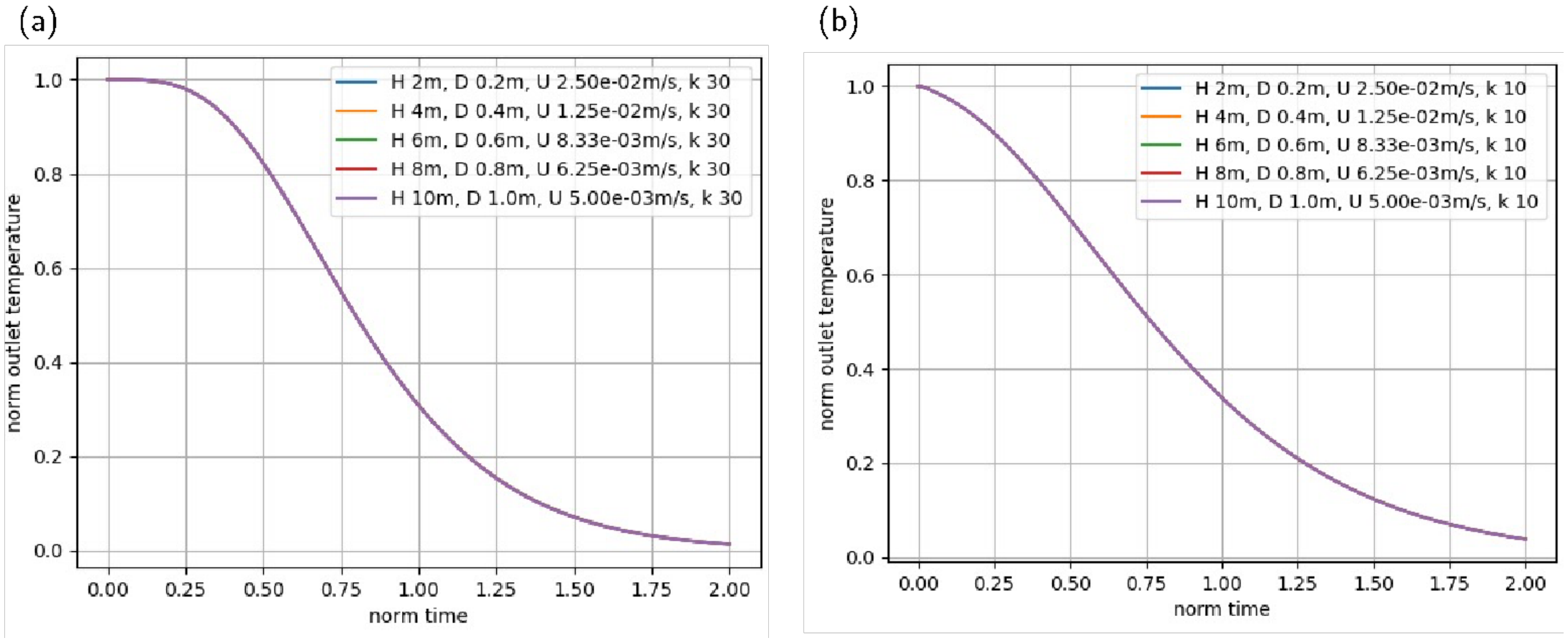

*Figure S2: validity of dimensional analysis (H=L)*

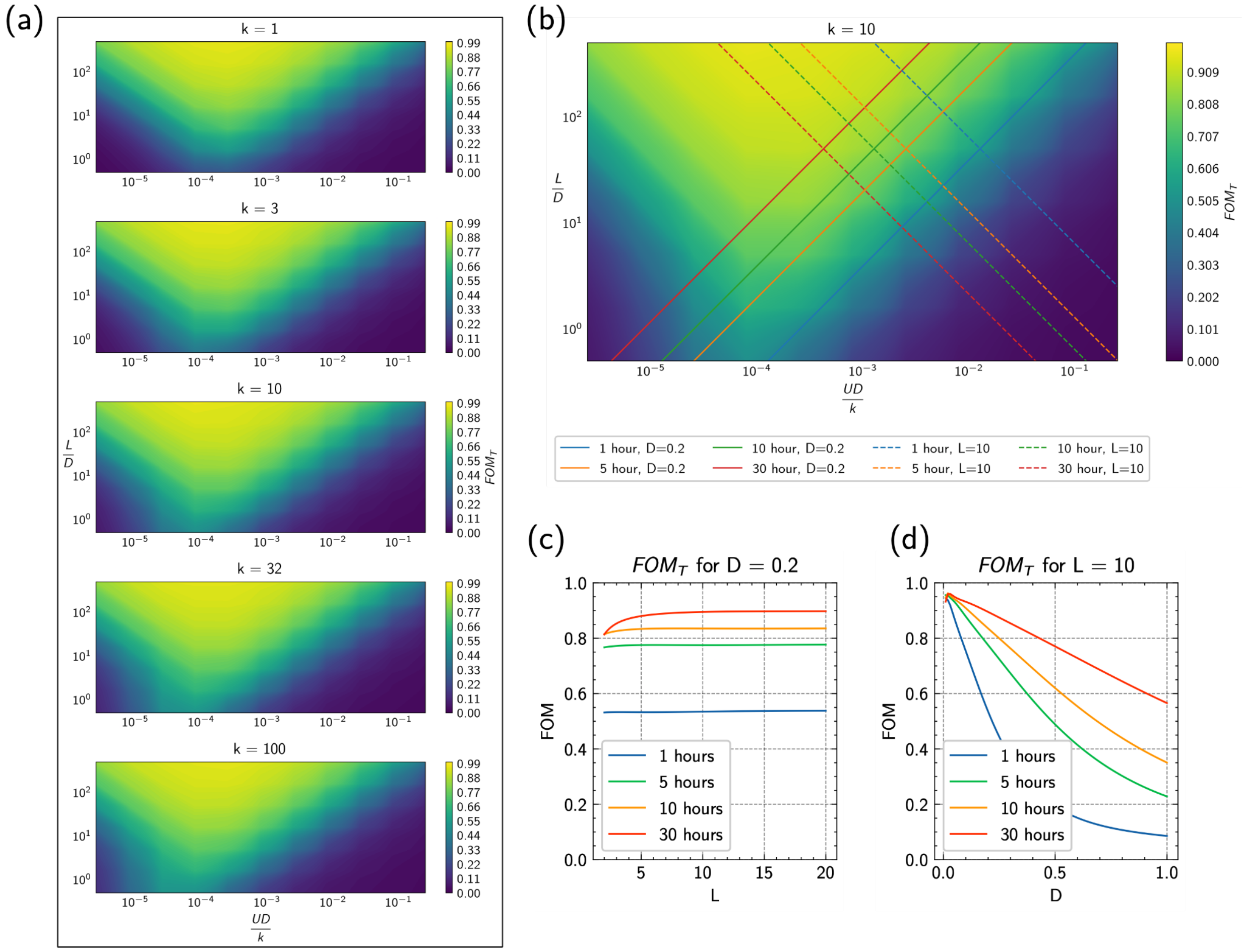

*Figure S3: Results from dimensional analysis showing the complete parameter space of the channel-embedded thermal energy storage system. (a) Sweeps of the parameter space for $\frac{L}{D}$ vs. $\frac{UD}{k}$ for various values of $k$. (b) A single slice of the 3D parameter space, for $k = 10\,W/mK$, demonstrating how the parameter space could be sampled for constant channel spacing (i.e. graphite diameter, solid lines) or constant channel length (dashed lines). (c) Results of sampling the parameter space given D=0.2m and varying the channel length L, and (d) sampling the parameter space given L=10m and varying the channel spacing D, for different storage durations. For (c) and (d), kernel ridge regression interpolation was used to obtain finer-grain results.*

Now that we have verified the dimensional analysis, we can map the design space by sweeping over the grouped parameters. Figure S3(a) shows $FOM_T$ vs $\frac{L}{D}$ and $\frac{UD}{k}$ for various values of $k$ (reproduced from main text).

We next conduct a deeper analysis of the $k = 10$ case. We probe the design space for constant diameter $D$ (solid lines) or constant flow length $L$ (dashed lines), for a given storage duration $\tau$, as shown in Figure S3(b). To see how $\tau$ defined in Equation 2 translates to a velocity, we can re-write it as

$$\tau = \frac{\rho_C V_C c_{p_C}}{\dot{m}_{Sn} c_{p_{Sn}}} = \frac{\rho_C L\pi(D^2 - D_{Sn}^2)c_{p_C}}{\rho_{Sn} U\pi D_{Sn}^2 c_{p_{Sn}}} \propto \frac{L}{U}$$

Then, we can plot how $FOM_T$ varies as these parameters change, as shown in Figure S3(c) and (d). For example, Figure S3(d) shows that for a given tin flow path length of 10m, decreasing the diameter significantly improves $FOM_T$.

## S4. Identifying optimal geometry for alternative configuration

In the main text we set the storage duration $\tau$ to 30 hours and the graphite thermal conductivity to 10 W/m/K to find the ideal diameter of the channel-embedded cylinder and tin flow path length. If instead we were interested in a longer storage duration of 100 hours and an alternative material with a thermal conductivity of 1 W/m/K, the optimization would be different, as shown below.

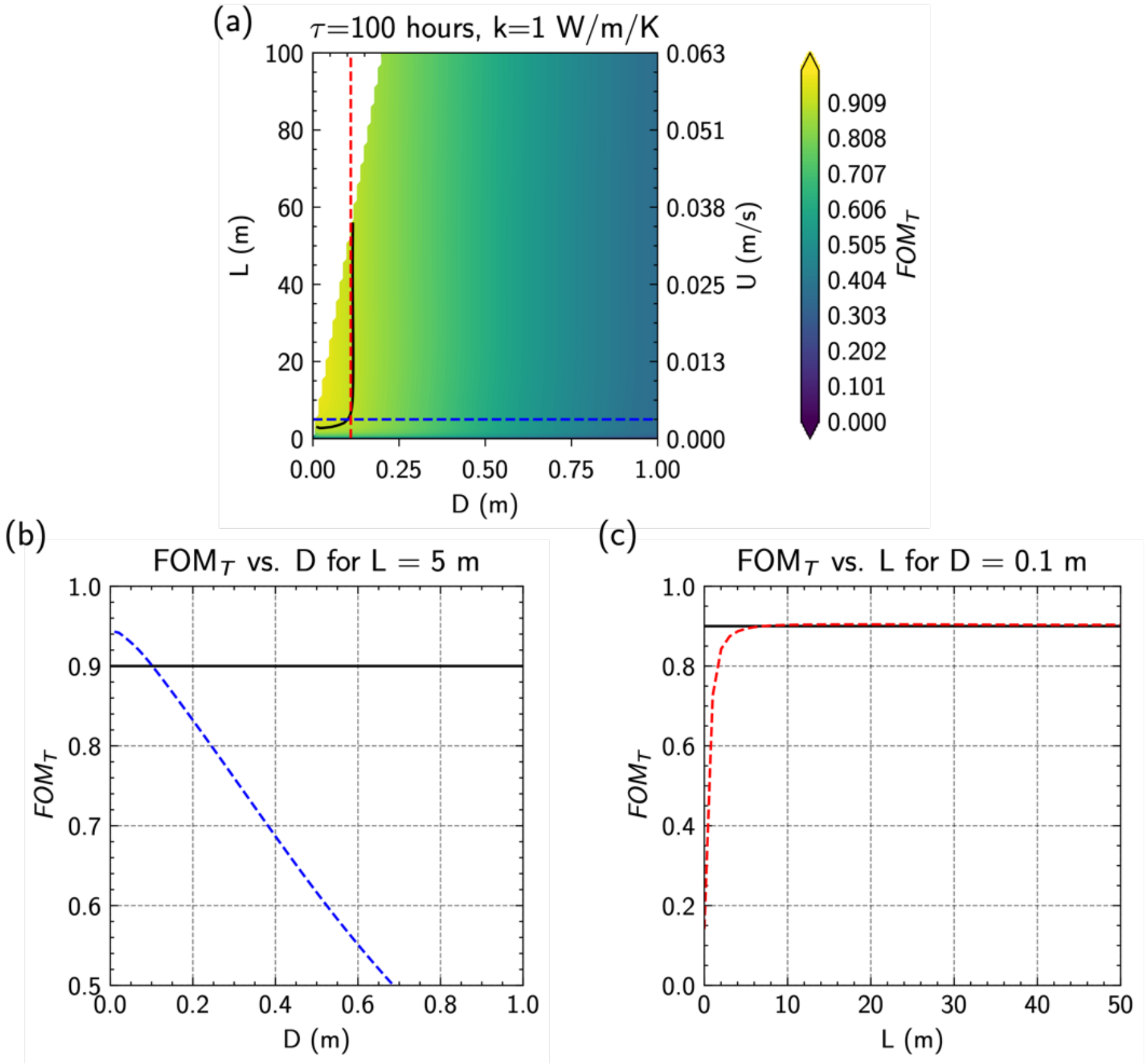

*Figure S4:*

## S5. Checking validity of laminar flow approximation

In certain flow cases, the Reynolds number of the flow may be higher than 4000. We are interested in determining if this impacts the heat transfer. We first calculate the Reynolds number for liquid tin flowing through a 2cm pipe at 1m/s:

$$Re = \frac{\rho V D}{\mu} = \frac{6184\left[\frac{kg}{m^3}\right] \cdot 1.32\left[\frac{m}{s}\right] \cdot 0.02[m]}{0.0011\left[\frac{kg}{m \cdot s}\right]} \approx 150{,}000$$

We then calculate a Prandtl and Peclet number as

$$Pr = \frac{\nu}{\alpha} = \frac{c_p\mu}{k} = \frac{244 \cdot 0.0011}{60} = 0.0044$$
$$Pe = RePr = 663$$

Using correlations from Taler,[38] we find that

$Nu = 5 + 0.025Pe^{0.8} = 9.5$ (const surf temp)

$Nu = 4.82 + 0.0185Pe^{0.827} = 8.8$ (const heat flux)

Which suggests that the convective heat transfer coefficient in turbulent flow is approximately twice that of laminar flow. Although this then suggests that the convective contribution to the heat transfer resistance in turbulent flow is halved – however, the dominant resistance is still conduction through the graphite block, which is approximately 2 orders of magnitude greater. To prove this, we calculate the discharge profiles for both laminar and turbulent flow, for various discharge durations. As shown in Figure S5, these two flow conditions have very similar heat transfer performance.

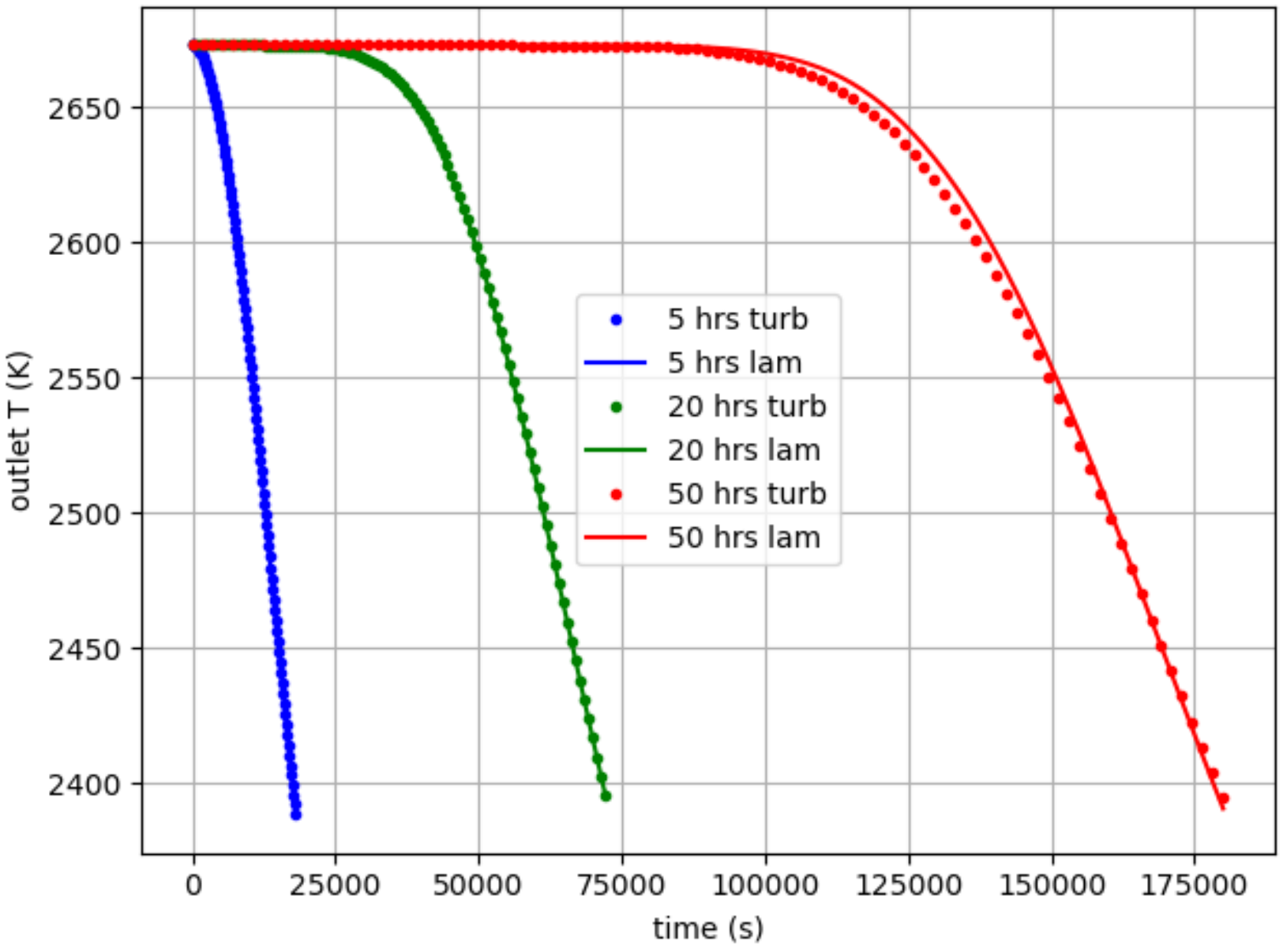

*Figure S5: validity of laminar approximation*

$$FOM_{P,charge} = \frac{\int_0^\tau \dot{m}(t)c_{p,Sn}\Delta T(t)\,dt}{E} = \frac{\int_0^\tau \dot{m}(t)c_{p,Sn}\Theta(t)\,dt}{E/\Delta T_{max}} = \frac{\int_0^1 \dot{m}^*(t^*)c_{p,Sn}\Theta^*(t^*)\,\tau dt^*}{E/\Delta T_{max}}$$

$$\frac{\int_0^1 \dot{m}^*(t^*)c_{p,Sn}\Theta^*(t^*)\,dt^*}{E/\tau/\Delta T_{max}} = \frac{\int_0^1 \dot{m}^*(t^*)c_{p,Sn}\Theta^*(t^*)\,dt^*}{P/\Delta T_{max}} = \frac{\int_0^1 \dot{m}^*(t^*)c_{p,Sn}\Theta^*(t^*)\,dt^*}{\dot{m}_{nom}c_{p_{Sn}}}$$

where $\dot{m}_{nom} = \frac{P_{max}}{c_{p_{Sn}}(500°C)} = \frac{E/\tau}{c_{p_{Sn}}(500°C)}$ and $\dot{m}^*(t^*)$ is flowrate as a function of non-dimensional time, such that $\dot{m}^*(t^*) = \dot{m}(t)$.

## *S6. Checking validity of porous media approximation*

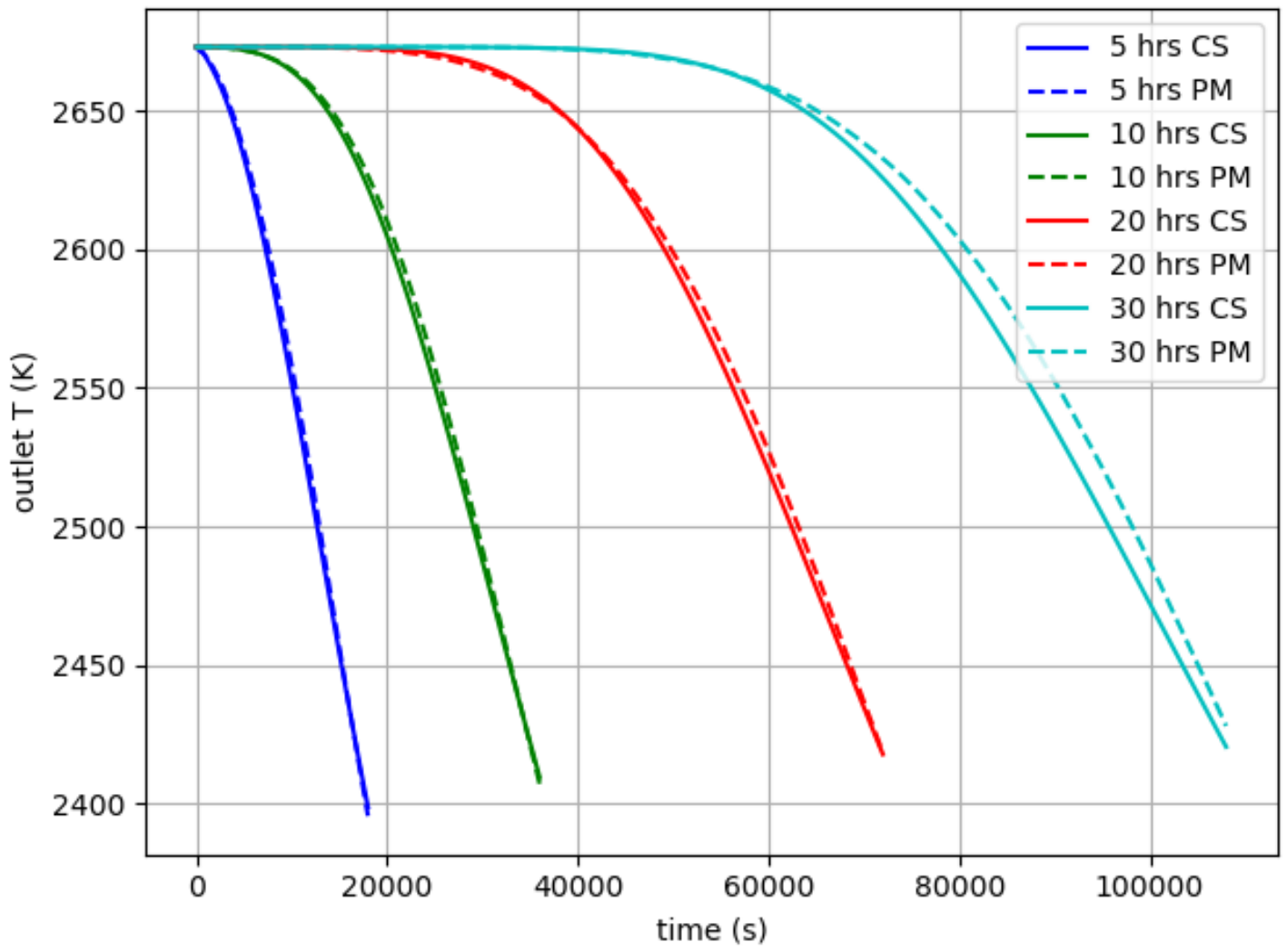

*Figure S6: validity of porous media approximation*

## S7. Evaluating options for accelerated charging

From our dimensional analysis, we found that the limiting heat transfer resistance is conduction through the storage material, especially if its thermal conductivity is low. Therefore, we focus on improving heat conduction. Given that conduction heat transfer can be approximated by $q" = \frac{k\Delta T}{L}$, we could either reduce the conduction length, or increase the $\Delta T$.

We consider 4 additional options for maximizing $FOM_{P,charge}$. The first two options rely on reducing conduction length, while the last two rely on increasing the $\Delta T$. For option 1, we could reducing the conduction distance in the storage material, either by (a) scaling down both the storage block diameter and the HTF channel diameter, or (b) keeping the HTF channel diameter constant while reducing the storage block diameter, For option 2, we similarly reduce the conduction distance in the storage material, but this time by relying on radiation to transfer heat between the HTF channel and a cylindrical shell of storage material (with the same volume as the original cylindrical block). For option 3, we could increase the inlet temperature of the HTF above the highest operating temperature of the storage material, to accelerate the heat conduction into the storage material by increasing the temperature difference. For option 4, like the methodology presented in the main text, we could increase the flowrate, but in contrast, always have the flowrate at a high value. This increases the power output throughout the charging process, so requires additional heating power, but may be more effective. We analyze all 4 options in the context of the TEGS system.

No new methodology is introduced for option (1) or (4). For option (2), we note that the radiative resistance between the tin pipe and the graphite block is potentially much smaller than the conduction resistance through the graphite block, due to the high temperatures of the system. By adding a gap between the tin tube and the cylindrical graphite shell, we can reduce the thickness of the shell while keeping the volume of graphite constant, reducing conduction resistance. For option (3), we keep the amount of thermal energy in the blocks the same, but increase the maximum tin inlet temperature, which allows greater charging power from Equation 1. If the amount of energy input exceeds the energy capacity of the blocks, the charging is considered complete and input power is set to 0.

To summarize, option 1 involves resizing the storage system, option 2 involves adding a spacing to exploit low radiative resistance, option 3 involves supercharging the tin inlet temperature above 2400ºC, and option 4 involves increasing the charging power.

For option 1, we re-do the geometry optimization from Section 4.1 given shorter storage durations, namely 5 and 3 hours, as shown in Figure S7.

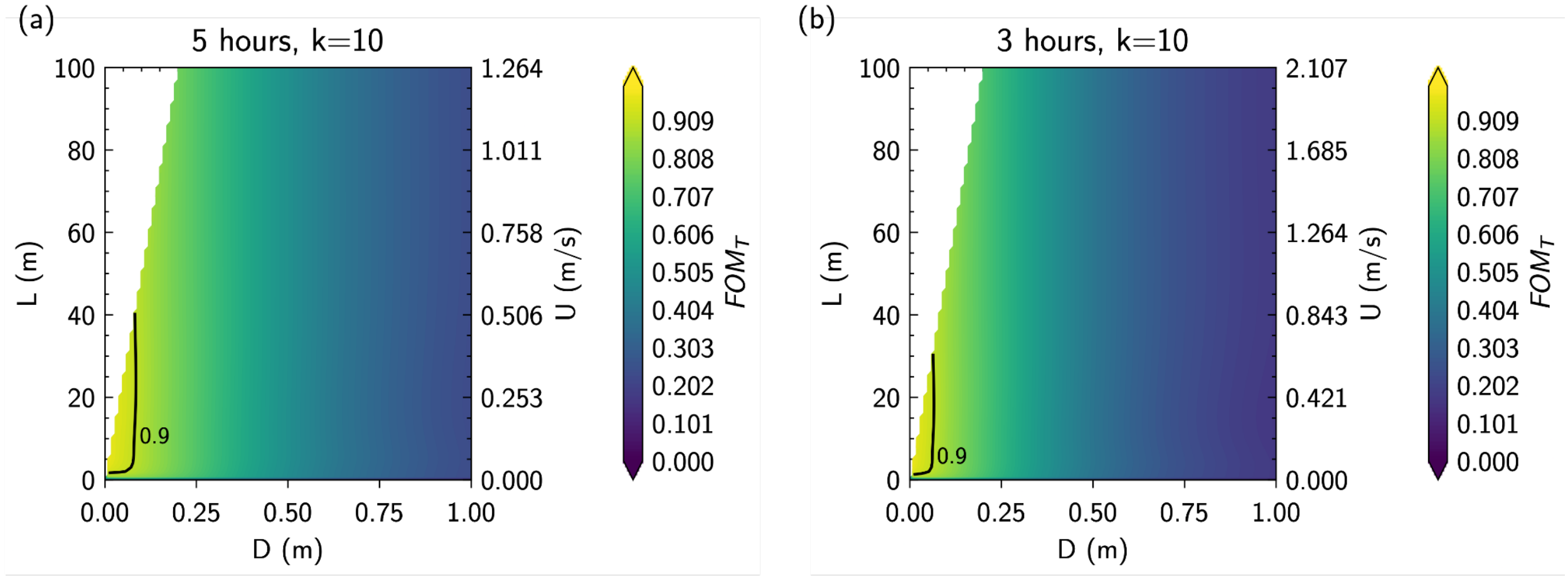

*Figure S7: Geometric parameter space for storage durations of (a) 5 hours and (b) 3 hours, similar Figure 4.*

As shown, to achieve 90% $FOM_T$, this would require a block diameter of ~8cm for 5 hours and ~6cm for 3 hours. These correspond to tin tube diameters of 8mm and 6mm, respectively. This would drastically increase the pressure drop, increase the velocity corresponding to an increased likelihood of cavitation, and potentially introduce surface tension effects at these small diameters. Therefore, this resizing is impractical to implement.

Detailed analysis of option 1(b) is not presented here as it would require a separate dimensional analysis since the fraction of $\frac{D_{Sn}}{D_C}$ is not assumed to be constant. Generally, from the analysis above, it is expected that increasing the channel diameter without increasing the graphite diameter would improve FOM. However, it would increase cost as more tin would be required per volume of graphite, and it would reduce energy density, as more overall volume would be required to have the same volume of graphite. Therefore, this option is not practical.

The second option for fast charging is to add a spacing between the tin tube and graphite block and exploit the low radiative resistance at these high temperatures. The radiative resistance and conduction resistance within the graphite block can be written as

$$R_{rad} = \frac{1}{\epsilon\sigma(T_1^2 + T_2^2)(T_1 + T_2)2\pi r_p L F_{12}}$$

$$R_{cond} = \frac{ln\left(\frac{r_o}{r_i}\right)}{2\pi L k_C}$$

Note that the convective resistance of the tin flow and the conductive resistance of the tin pipe are neglected here. The parameters of the optimized geometry are $r_p = r_i = 0.01m$, $r_o = 0.2m$, and $L = 10m$. We can investigate the impact of adding a gap between the tin flow pipe and the graphite block, such that $r_i > r_p$, by calculating the resistances. Note that $r_o$ is adjusted as $r_i$ is increased to keep the graphite block volume $V = \pi(r_o^2 - r_i^2)L$ constant, such that the thickness $r_o - r_i$ decreases for larger $r_i$, as shown in Figure S8(a). $F_{12}$ is calculated using the equal finite concentric cylinders formula found in Brockmann.[39] We assume $k_C = 10\frac{W}{mK}$, $T_1 = 2400°C$, and $T_2 = 1900°C$.

Figure S8(b) shows the radiative resistance, conductive resistance, and total resistance when sweeping over different values of gap size. As seen, adding a small gap increases the total resistance, but this effect is mitigated after a gap of ~3mm. Adding a larger gap reduces the total resistance, primarily due to the lower conductive resistance created since $\frac{r_o}{r_i}$ is lower at high $r_i$. Adding a gap of 5cm reduces the overall resistance by approximately half.

While the resistance analysis is a useful guide, it assumes steady state. To investigate the transient performance, we model the full 3D geometry with surface-to-surface radiation and various gap sizes. These results are shown in Figure S8(c). As seen, the peak performance occurs at around 4cm gap. Below this value, the conduction resistance is too high, and above this value, too much heat leaks out of the top and bottom of the graphite block. Note that this effect could be mitigated by adding a radiation shield to reflect the lost heat.

The tradeoff with adding this gap is that the storage system takes up more volume, which reduces the energy/power density and increases the amount of insulation required. Figure S8(d) quantifies this effect, showing the percent increase in volume and surface area above the nominal for various gap sizes. As seen, the slight benefit to adding a radiative gap is outweighed by the large increase in surface area.

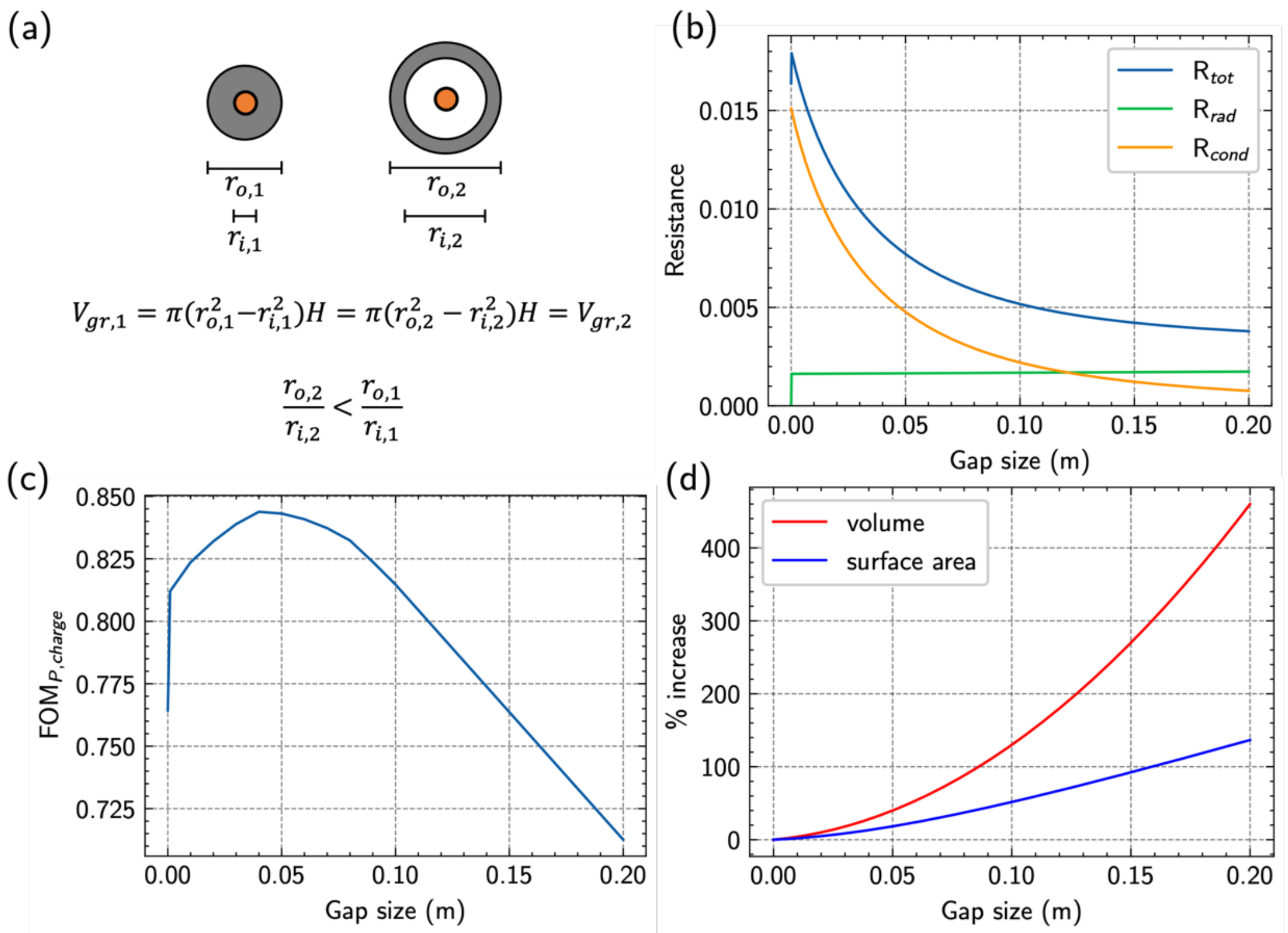

*Figure S8: (a) Analysis of contributions to heat transfer resistance, considering both radiation between the tin channels and the graphite block and the conduction through the graphite block. (b) The effect of the size of the gap between the tin channel and the graphite block on the charging FOM. (c) The tradeoff of increasing the gap size, namely the increase in volume and outer surface area of the storage section and therefore increase in insulation costs.*

The third option is to supercharge the inlet tin to above 2400ºC. This increases the temperature difference between tin and graphite and therefore increases the heat flux. The charging process was simulated for various tin inlet temperatures, as shown in Figure S9.

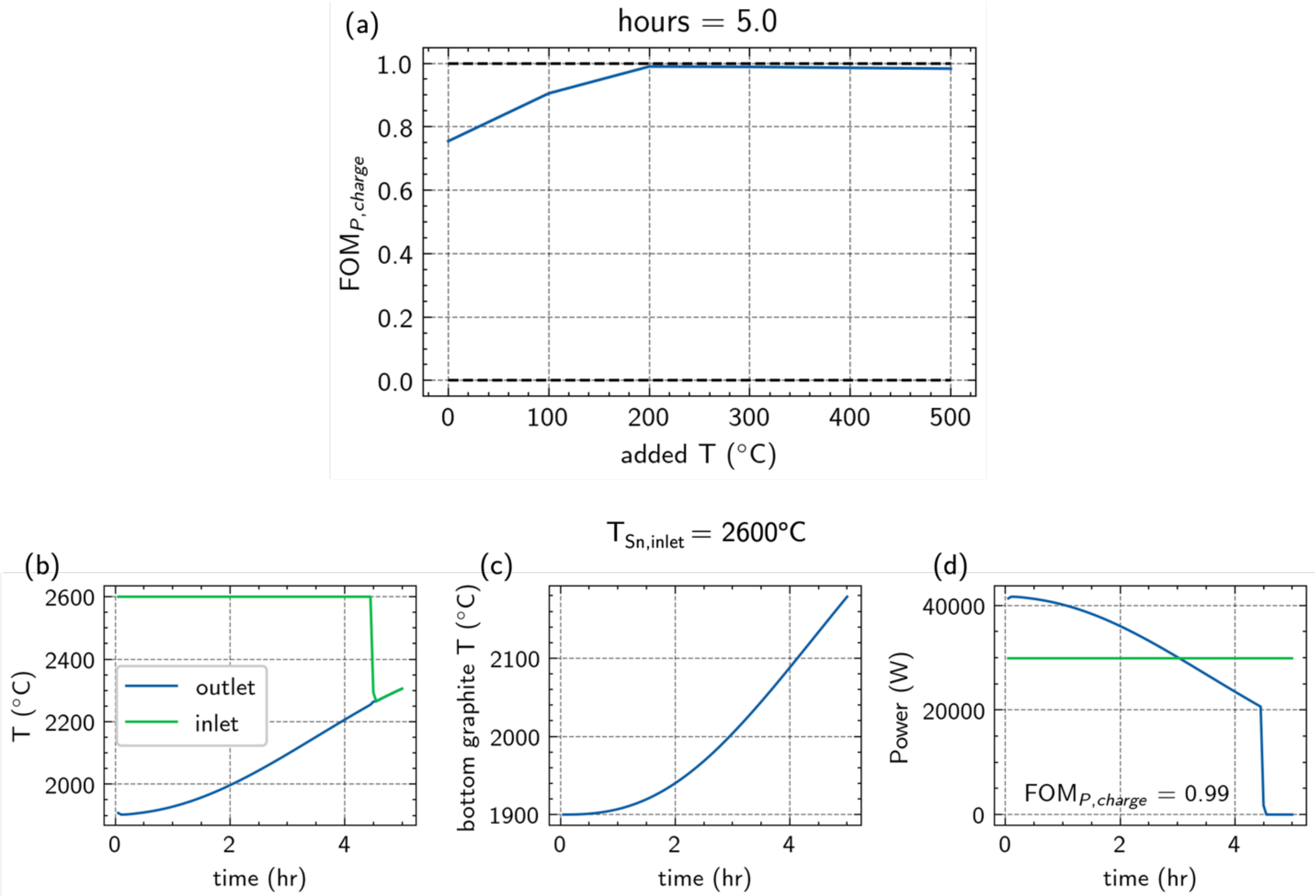

*Figure S9: (a) Effect on charging FOM of superheating the tin above 2400ºC during the charging process. Then, for an added temperature of 200K, (b) outlet vs. inlet temperature of tin to the storage media, showing charging completed after 4.25 hours indicated by inlet temperature dropping to outlet temperature, (c) lowest temperature of the storage medium as a proxy of charging completeness, and (d) power input as a function of time, with the average power required to fully charge shown in green and the actual power input in blue, and the corresponding charging FOM shown.*

In this case, we keep the thermal energy capacity of the blocks constant, at $E = mc_{p,C}(500°C)$, but allow the input power to increase to $P = \dot{m}c_{p,Sn}(500°C + T_{added})$. As seen in Figure S9(a), setting $T_{added}$ to be $200°C$ allows $FOM_{P,charge}$ to reach ~1. Specifically, as seen in Figure S9(b), the block is fully charged in around 4.25 hours, after which the tin inlet temperature is dropped to the outlet temperature to prevent further charging. In this case "fully charged" means the average block temperature is 2400ºC, however after 4.25 hours there is still temperature variation in the block, as seen in Figure S9(c), which shows the temperature at the bottom right of the block, which is the lowest temperature. Therefore, additional time is required to reach thermal equilibrium at 2400ºC. Another potential problem with this approach is the high vapor pressure of tin at temperatures above 2400ºC and boiling of tin above 2600ºC. This could lead to loss of tin in the flow paths and condensation in other areas. It could also lead to cavitation at high velocities.

The fourth option is to increase the flowrate and therefore charging power through the entire charging process. We evaluate the effectiveness of this methodology in the context of the TEGS system. For this analysis, we consider a graphite block of the same dimensions as the previous subsection (length $L = 10m$, diameter $D_C = 0.2m$, tin channel diameter $D_{Sn} = 0.02m$, determined from dimensional analysis). For the accelerated charging analysis, we first consider a short charging duration of $\tau = 5$ hours, which corresponds to, for example, the amount of time abundant solar might be available during the day. The nominal flowrate for such a system would be $\dot{m}_{nom} = \frac{E/5h}{c_{p,Sn}(500°C)}$. To accelerate charging, we define a new maximum flowrate $\dot{m}_{max}$ and a flowrate multiplier $f = \frac{\dot{m}_{max}}{\dot{m}_{nom}}$. In contrast to the constant power discharging technique, instead of ramping up the flowrate with time, the flowrate is kept at the maximum value $\dot{m}_{max}$ to maximize power input. The results for the given system, with 5 hours of charging duration and various flowrate multipliers, are shown in Figure S10.

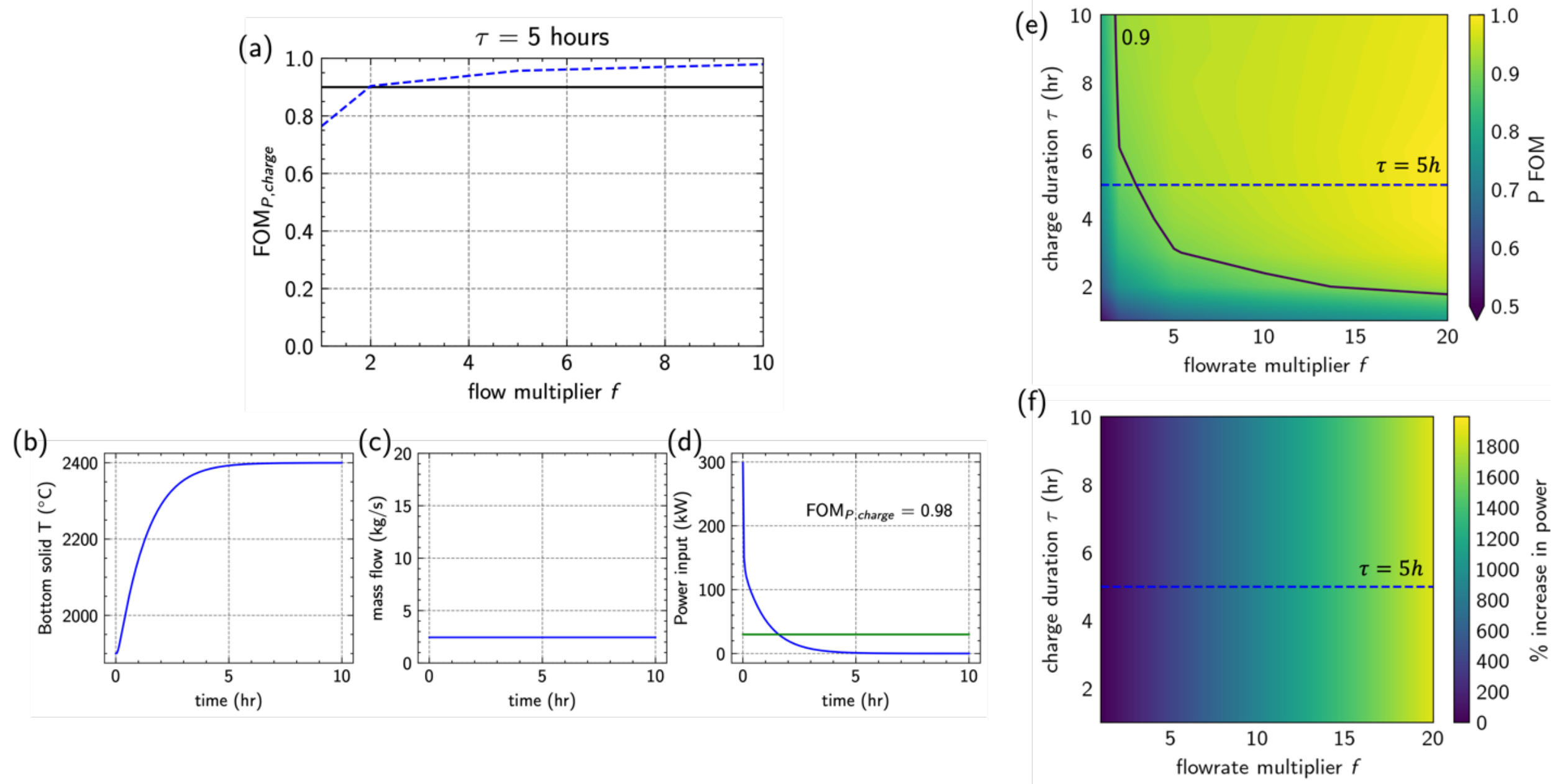

*Figure S10: (a) Increase in $FOM_{P,charge}$ as the flowrate increases, for a charging duration of 5 hours. Then, considering a flowrate multiplier of 10, (b) the lowest temperature of the storage medium as a function of time, as a proxy for charging completeness, (c) mass flowrate of tin as a function of time, showing it is constant at 10x the nominal, (d) power input as a function of time, where the green line is the average power input required to fully charge the system, and the blue line is the actual power input. (e) Conducting the analysis for various other charging durations and flowrate factors, with the solid black line indicating an isocontour of FOM=0.9 and the blue dashed line indicating a charging duration of 5 hours. (f) Increase in charging power associated with each flowrate multiplier, where the % increase is relative to the f=1 case for each duration. This increase in charging power will increase the charging infrastructure cost.*

As seen in Figure S10(a), increasing the flowrate from 1x to 10x the nominal improves $FOM_{P,charge}$ from 75% to 98%. The specific response of the thermal storage system to 10x the nominal flowrate is shown in Figure S10(b)-(d). Figure S10(b) shows the temperature of the bottom of the graphite block (near the tin outlet), as an indication of the completeness of charging since the bottom charges the slowest. As seen, within 5 hours the temperature almost approaches the maximum value of 2400ºC. However, as seen in Figure S10(d), this requires a higher input power at the beginning, increasing the cost of heating infrastructure, specifically by a factor of 10 (from 30 to 300 kW).

Next, we consider various charging durations from 1 to 10 hours, as shown in Figure S10(e). Similar to discharge power uniformity, the accelerated charging is easier at longer durations since their charging FOMs are already high even with no modifications (at $f = 1$). We note that for all durations >3 hours, we are able to nearly fully charge the system (>90%) with flowrate multipliers of 5 or lower.

As mentioned previously, the increase in performance comes at a cost of increasing charging power, as shown in Figure S10(f). This may not be significant if the charging infrastructure is cheap. While increasing HTF flowrate to establish a near-constant HTF temperature within the channel is effective, it increases pressure drop and requires additional pumps and heating power, so there is again a tradeoff between performance and cost.

While the options all enable increases in $FOM_{P,charge}$, they were deemed impractical for various reasons, which are summarized here. For option 1(a), the smaller tin tube diameter introduces issues with higher pressure drop, surface tension, and cavitation. For option 1(b), the smaller graphite diameter means less volume of storage material, which means additional volume is required which increases insulation costs and reduces energy density of the battery. For option 2, the cylindrical shell geometry similarly requires additional area for insulation (and lower energy density) as well as increasing the heat loss out of the system due to the gap between the tin channel and the block. For option 3, the higher tin temperature introduces issues with sublimation and boiling. For option 4, the increase in charging power cost is impractical.

## *S8. GIFs of temperature profile vs. time for full-scale simulation*

GIFs of temperature profile vs. time for 10x10 grid, with 1 path vs. 10 parallel paths available here:
https://github.com/shomikverma/graphite-discharge/blob/main/plots/GIF_discharge_animation.gif

## *S9. Charge/discharge dynamics*

In this paper, we have discussed charge and discharge cycles separately, but another important consideration is the charge-discharge transition. For example, if the cost of electricity suddenly drops while discharging, the system may want to immediately switch to charging. To optimize this process, we introduce the concept of flow reversal, to ensure the outlet temperature of the storage blocks is optimized. For example, for uni-directional flow, halfway through discharge, the temperature of the end of the block is still relatively high. Thus, there is limited charging potential. In contrast, if the flow is reversed, halfway through discharge, the top of the block is likely cold, allowing full charging capability.

To analyze this phenomenon, we investigate each case (unidirectional vs. reversed) separately. For both cases, we consider a block designed for 19 hours of storage and model several cycles of 9.6 hours of discharging and 2.4 hours of charging. We use Equation 2 to find the power required for each of discharge or charge. To ensure constant power during discharge and charge, we use Equation 3 to vary the flowrate, using a maximum flowrate factor of 10. During discharge, we use an inlet tin temperature of 1900°C, while during charge we use an inlet tin temperature of 2400°C.

For uni-directional flow, we keep the top of the block as the inlet and the bottom as the outlet. For flow reversal, we use the top of the block as the inlet for discharging, and the bottom of the block as inlet for discharging. Figure S11 shows the results for 72 hours of simulation, for both uni-directional ((a)-(c)) and reversed ((d)-(f)) flow.

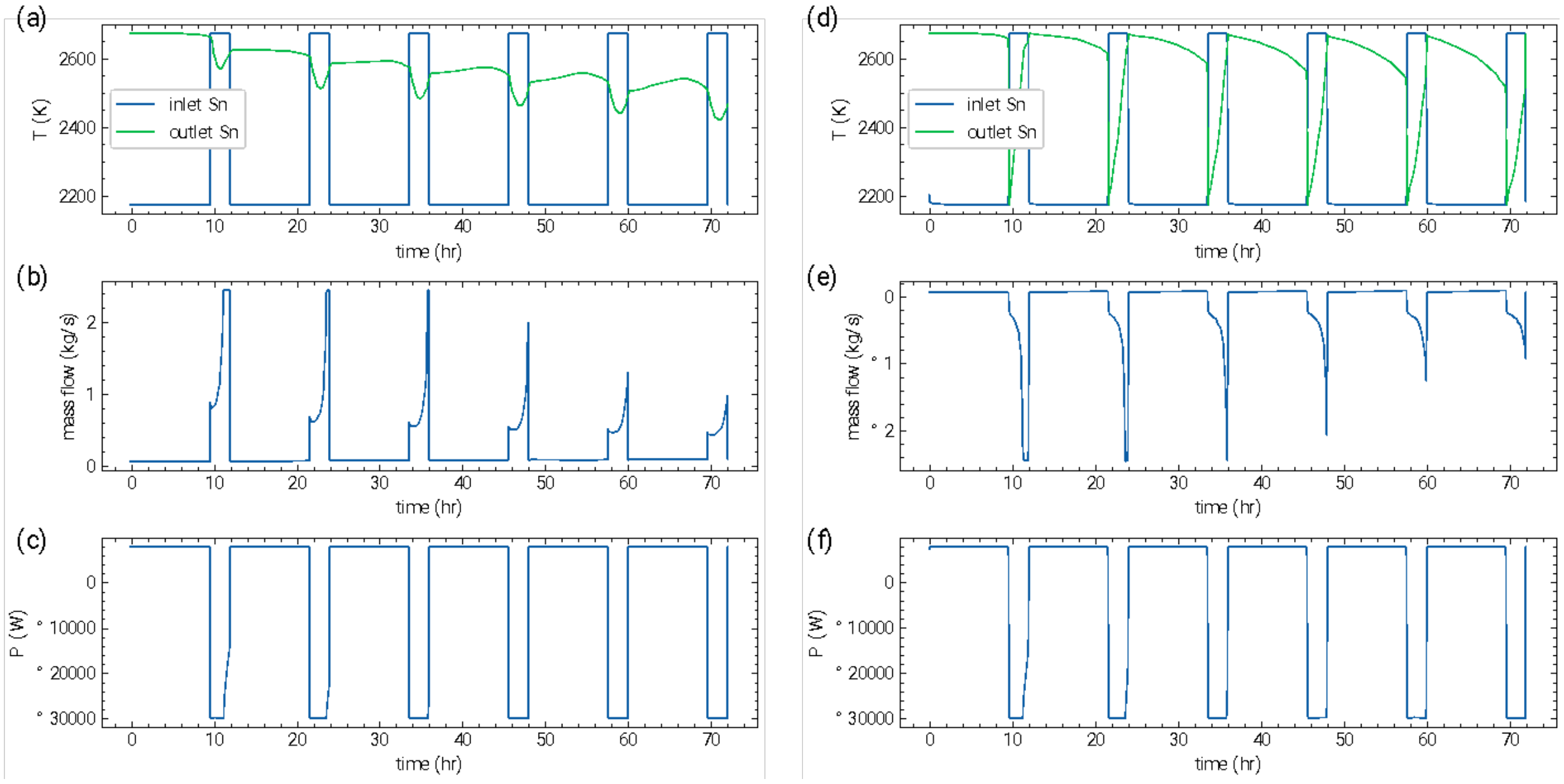

*Figure S11: Plots of temperature, massflow, and power for (a)-(c) unidirectional and (d)-(f) reversed flow.*

As seen, flow reversal allows quicker transition from hot outlet temperature to cold, and vice versa. Due to the flexibility in flowrate, the uni-directional case can achieve similar thermal power output/input as the reversal case. If flowrate was set to be constant, this would not be the case. Further, when converting thermal to electrical power, TPV power density and efficiency must be considered. Because flow reversal enables higher temperatures, the discharge power will be higher for a given TPV area.

Other arguments for flow reversal include minimization of exergy destruction and limiting thermal cycling of the graphite blocks. In the unidirectional case, during charging hot tin is put in contact with cold graphite, destroying more exergy than if hot tin was put in contact with hot graphite. In the flow reversal case, one side of the graphite remains hot while the other remains cold, reducing the number of thermal cycles the graphite sees.

GIFs of unidirectional vs. reversed flow available here:
https://github.com/shomikverma/graphite-discharge/blob/main/plots/GIF_charge_discharge_animation.gif